\documentclass[letter,11pt]{article}
 
\usepackage{jheppub}

\usepackage{graphicx}
\usepackage{dcolumn}
\usepackage{bm}
\usepackage{float}
\usepackage{hyperref}
\usepackage{dsfont}
\usepackage{slashed}
\usepackage{color}
\usepackage{amsmath}
\usepackage[section]{placeins}
\usepackage{braket}
\usepackage{upgreek}
\usepackage[bottom]{footmisc}
\usepackage[normalem]{ulem}

\newcommand{\bea}{\begin{eqnarray}}
\newcommand{\eea}{\end{eqnarray}}

\def\cs{{\rm cs}}
\def\ucs{{\rm ucs}}
\def\d{{\rm D}}

\def\med{{\rm med}}

\def\s{{\rm s}}
\def\dr{{\rm d}}

\def\bs{\boldsymbol} 
\def\bfk{{\bs k}}
\def\bfq{{\bs q}}
\def\bfu{{\bs u}}
\def\bfr{{\bs r}}

\def\bfb{{\bs b}}
\def\bfp{{\bs p}}
\def\bfl{{\bs l}}
\def\bfP{{\bs P}}
\def\bfQ{{\bs Q}}

\title{Towards factorization with emergent scales  for jets in dense media}

\author{Balbeer Singh and}
\author{Varun Vaidya}
\affiliation{Department of Physics, \\ 
University of South Dakota}

\emailAdd{Balbeer.Singh@usd.edu}
\emailAdd{Varun.Vaidya@usd.edu}

\date{\today}
\preprint{}

\abstract{Employing the recently developed open quantum system Effective Field Theory framework, we investigate jet production and evolution in a dense nuclear medium in electron-ion/heavy-ion collisions. We confirm that the frequent monitoring of the jet by the medium leads to the emergence of a perturbative transverse momentum scale, often referred to as the saturation scale that necessitates further factorization to completely isolate the non-perturbative physics of the medium. A part of this goal is achieved in this paper by providing an operator definition for the  broadening probability of a gluon in the medium within the Markovian approximations. We show that this distribution is (semi)universal; it depends on the angular measurement on the jet and probes both the large and small $x$ dynamics of the medium. We further elucidate all other contributions to non-perturbative physics suggesting that the parameterization of non-perturbative physics is more complex than previously assumed and outline steps required for a complete factorization of the jet production cross section.}

\begin{document}

\maketitle

\section{Introduction}
One of the most natural and fundamental tool for accessing the microscopic structure of strongly coupled phases of Quantum Chromodynamics(QCD) in the high energy nuclear collision experiments are highly energy jets.  Jets are collimated sprays of particles, i.e., quarks and gluons which are created in high energy particle as well as nuclear collision experiments.  
As was anticipated in Ref.~\cite{Bjorken:1982tu}, due to the energy loss of energetic color charge parton while traversing the medium, the suppression of these jets or high transverse momentum ($p_T$) particles, may signal the formation of quark-gluon plasma (QGP) in hadron-hadron collision experiments.
Eventually, the remarkable phenomena of ``jet quenching'' was observed first in Au-Au collision at the Relativistic Heavy Ion Collider (RHIC) and later at the Large Hadron Collider (LHC). The discovery of this deconfined state of QCD matter has prompted extensive theoretical and experimental efforts to understand the properties of QGP~\cite{BRAHMS:2004adc,PHOBOS:2004zne,STAR:2005gfr,PHENIX:2004vcz,ATLAS:2010isq,
ALICE:2010yje,
CMS:2011iwn,ATLAS:2018gwx,CMS:2021vui,ALICE:2023waz,Connors:2017ptx,Wiedemann:2009sh}. For recent studies see Refs.~\cite{Ke:2024emw,Chien:2024uax,Budhraja:2023rgo,Barata:2024bmx,Mehtar-Tani:2024mvl,Barata:2024ieg,Li:2024pfi}



While the understanding of production cross-section of jets in systems like proton-proton (pp) and electron-proton ($ep$) collisions, in which the jet evolves in vacumm, has acquired unprecedented quantitative precision in recent years (See Refs.~\cite{Larkoski:2017jix,Asquith:2018igt,Marzani:2019hun} for latest reviews), similar quantitative rigor has not yet been achieved for jet quenching theory in HICs.
The tremendous progress in simpler systems such as $ep$ and $pp$ has  been made possible by utilizing the paradigm of factorization ~\cite{Collins:1989gx} that allows for the separation of various dynamics  in terms of distinct functions and also enable predictions when the target is strongly coupled. Factorization theorems, which can be proved systematically by using the framework of Effective Field Theory (EFT), lead to a clean demarcation of physics at widely separated scales allowing us to isolate the non-perturbative physics through well-defined objects such as parton distribution functions (PDFs), and extracting them from a set of reference processes and lattice data. These functions carry universal information about the partonic structure of the strongly coupled system. Moreover, the universality of these functions also allows us to make predictions for new jet substructure observables. It is therefore desirable to have an equivalent framework for jet production and evolution in heavy-ion collision environment.


Implementation of tools of factorization and resummation for the quenching of jets in HICs has been hitherto difficult mainly due to the plethora of new effects that appear through the interactions of fast moving color charge parton in the jet and the medium. In particular, this includes the Landau-Pomeranchuk-Migdal (LPM) effect \cite{Landau:1953um,Migdal:1956tc}, which is a quantum interference effect of multiple scattering centers in the medium that causes energetic color charge partons to lose energy to the plasma.
This coherent effect of multiple scatterings was first understood in the early 1990s~\cite{Gyulassy:1993hr,Wang:1994fx,Baier:1994bd,Baier:1996kr,Baier:1996sk,Zakharov:1996fv,Zakharov:1997uu,Gyulassy:2000er,Wiedemann:2000za,Guo:2000nz,Wang:2001ifa,Arnold:2002ja,Arnold:2002zm} based on which phenomenological models were developed to understand the qualitative features of the experimental data~\cite{Salgado:2003gb,Liu:2006ug,Qin:2007rn,Armesto:2011ht}. Moreover, due to the multi-partonic nature of the jet, another phenomena known as color (de)coherence dynamics driven by the interference patterns between multiple fast moving partons also needs to considered~\cite{Mehtar-Tani:2010ebp,Mehtar-Tani:2012mfa,Casalderrey-Solana:2011ule, Casalderrey-Solana:2012evi,antenna_dense}. These interferences, contributing to the energy loss of the jet, depends on the resolution power of the QGP.  


As a step towards the goal of utilizing the tools of factorization, a systematically improvable EFT framework to understand jet production in dense nuclear media accounting for all the above mentioned effects was formulated for the first time in Ref~\cite{Mehtar-Tani:2024smp}. For substructure observable that applies this formalism see Ref~\cite{Singh:2024vwb}. The main result of this work was a factorization formula for the jet production cross section in HICs by utilizing Soft Collinear Effective Theory(SCET) and its Glauber extension \cite{Bauer:2000yr, Bauer:2002nz, Rothstein:2016bsq}.
The explicit computations in this work were limited to the case of a dilute medium which involves only a single interaction of the jet with the medium. 
In this paper we implement the factorization formula derived in Ref~\cite{Mehtar-Tani:2024smp}  to study the case of jet production and evolution in a dense medium taking into account multiple interactions of fast moving color charge parton in the jet with the medium. Within the EFT framework, we quantitatively prove the emergence of a perturbative saturation scale using an explicit calculation in the Markovian limit. While such a scale has been observed in earlier calculations~\cite{Casalderrey-Solana:2011ule}, we show that the presence of this scale dictates that the jet probes both large and small $x$ dynamics of the medium. This is encapsulated in a broadening probability distribution which we write in terms of two non-perturbative functions. We show  that the current parametrization of jet broadening in terms of the jet quenching parameter($\hat q$) is an approximation which can be recovered from our result in a certain limit.  We also uncover other sources of non-perturbative physics dictated by the virtuality of vacuum and medium induced radiation. Therefore to fully isolate the non-perturbative physics for jet quenching is requires further work and we outline the steps needed for the same.

This paper is organized as follows. In Section \ref{sec:FactR} we review the factorization formula derived in \cite{Mehtar-Tani:2024smp} and use it to obtain the resummed cross section for a dilute medium in Section \ref{sec:DilM}. We then move on to the one loop computation for jet broadening in a dense medium in Section \ref{sec:DenseM}. Based on this calculation we discuss the emergent saturation scale  $Q_{\med}$ in Section \ref{sec:eScale} and its implications for non-perturbative physics. In Section \ref{sec:Match2} we look at the ingredients needed to perform a matching from the scale $Q_{\med}$ to $m_D$ and derive an Operator definition for the broadening probability of a gluon in the medium that reflects this scale separation. Finally, we discuss the implications of the results derived in this paper and outline further calculations needed in order to realize the matching procedure for a full factorization of all non-perturbative effects in \ref{sec:Con}.
\section{A review of factorization}
\label{sec:FactR}
In this section, we briefly review the factorization formula for jet production in the medium which was developed and presented in Ref.~\cite{Mehtar-Tani:2024smp}. 
For jets, the relevant production cross section is a histogram of number of jets based on their transverse momentum $p_T$ and rapidity $\eta$, for a fixed jet radius $R$.  We work in the narrow-jet $R\ll 1$ limit, which has been studied in $pp$ collisions and is a good approximation even for relatively large values of $R$~\cite{Dasgupta:2014yra,Kaufmann:2015hma,Kang:2016mcy,Dai:2016hzf,vanBeekveld:2024jnx,Lee:2024icn}. 
While traversing the medium, jets experience additional scales such as medium's length $L$, mean free path of the jet $\lambda_{\rm mfp}$, temperature $T$, Debye mass $m_D\sim gT$, where $g$ is the QCD coupling  that are manifestly introduced by the medium. 
Moreover, the medium dependence of the jet observables is mainly encoded in a single emergent scale, which is associated with the effective transverse momentum gained by fast moving color charge parton, $Q^2_{\rm med} \equiv \langle k_\perp^2\rangle \sim \hat q L$, where $\hat q$ is the jet transport parameter \cite{Baier:1996sk,Casalderrey-Solana:2011ule}. 
Here, for illustration purposes, we consider a jet with $p_TR\sim 10-100$~GeV, and medium temperature $T\sim 0.5$~GeV, which is also achievable in current collider runs. Furthermore, we also set $\hat{q}\sim $1-2~\text{GeV}$^2/\text{fm}$ and medium length $L\sim $2-5~fm, hence, the intrinsic medium scale varies in the range $Q_{\rm med}\sim $1-3~GeV. It may be noted that these values are an adequate example of the hierarchy considered here 
\begin{equation}
p_T\gg p_T R \gg Q_{\rm med}\gtrsim T\sim \Lambda_{\rm QCD}\,.
\end{equation}
For the ranges of temperature considered here $g\sim \mathcal{O}(1)$, therefore, $m_D\sim T$ and are not separable and will be treated as same scales. At this stage, the value of $Q_{\med}$ is only a guide to derive our factorization formula, neverthless, we will obtain an estimate of this scale through an explicit calculation and verify the validity of this hierarchy. Based on this hierarchy a factorization formula for the cross section was derived in Ref.~\cite{Mehtar-Tani:2024smp} which reads as
\begin{equation}
 \frac{{\rm d}\sigma}{{\rm d}p_T{\rm d}\eta}= \sum_{i \in q, \bar q , g}\int_0^1 \frac{{\rm d}z}{z} \, H_i\left(\omega= \frac{\omega_J}{z},\mu\right) \, J_i(z, \omega_J,\mu)\,.
\label{eq:factI}
\end{equation}
Here, $H_i$ are the hard-scattering functions including also PDFs, which describe the production of a massless jet initiating parton $i$ with four-momentum $p^\mu$ and  $ \omega\equiv p^- $ denotes its large momentum component.  The functions $J_i$ describe the jet evolution where $z=\omega_J/\omega$ is the momentum fraction of the jet initiating parton $i$ that ends up in the measured jet. The momentum fraction $z$ is related to the jet $p_T$ and rapidity $\eta$ via $\omega_J = 2p_T \cosh \eta$. The jet function can be further factorized as
\begin{align}
\label{eq:J-S-fact}
 J_i(z,\omega_J,\mu) &=\int_{0}^{1} \dr z' \,\int_0^{+\infty}\dr \epsilon_L \, \delta(\omega_J'-\omega_J-\epsilon_L)\, \sum_{m}\prod_{j=2}^m\int \frac{\dr\Omega(n_j)}{4\pi} \mathcal{C}_{i\rightarrow m}\Big(\{\underline{n}\},z', \omega_J'= \frac{z'\omega_J}{z}, \mu,\mu_\cs\Big)\nonumber\\
&\qquad\, \otimes {\cal S}_{m} (\{\underline{n}\} , \epsilon_L,\mu_\cs)  \,, 
\end{align}
which is valid up to expansion $\mathcal{O}(Q_{\rm med}/p_T R)$ and $\epsilon_L$ is energy loss due to radiation out of the jet cone. Further, medium induced subjet function acquire the form
\bea \label{eq:soft-funct}
&&{\cal S}_{m}(\{\underline{n}\},\epsilon_L) \equiv    \text{Tr}\Big[U_m(n_m)...U_{1}(n_1)U_0(\bar n)\rho_M U^\dag_0(\bar n)U_1^\dag(n_1)...U_m^\dag(n_m)\mathcal{M}' \Big] \,,
\eea
where we use the shorthand notation $\{\underline{n}\}\equiv\{n_1,n_2,...,n_m\} $ for the direction of the collinear subjets inside the jet that also satisfies $n_i\cdot n_j \gg R^2$ for $i\neq j$   and the measurement $\mathcal{M}'= \Theta_{\text{alg}}\delta(\epsilon_L - \bar n \cdot p_{\rm out} )$. The Wilson line $U_0(\bar n)$, which ensures gauge invariance, describes an unresolved effective charge moving in the opposite direction. The coefficients $\mathcal{C}_{i\rightarrow m}$ in Eq.~(\ref{eq:J-S-fact}) describes the production of $m$ energetic partons inside the jet at pairwise angles larger than the color decoherence angle $\theta_c \sim 1/(Q_{\med}L)$ from the initiating parton $i$. The convolution in Eq.~\ref{eq:J-S-fact} involves angular integrations in the directions of the $m$ partons. This refactorization takes the same form as encountered in the context of non-global logarithms~\cite{Dasgupta:2001sh,Larkoski:2015zka} in Ref.~\cite{Becher:2015hka}. Below we discuss the set up in more detail.

\noindent 
{\bf Unresolved jet}: In this paper, we will focus on the case of an unresolved jet, i.e the medium cannot resolve multiple subjets so that the entire jet acts as a single coherent color source for cs radiation. This is valid in the limit when $R \leq \theta_c$. In this case, from Eq.~\ref{eq:J-S-fact} the jet function acquires the form 
\begin{align}
\label{eq:SingleS}
 J_i(z,\omega_J,\mu) &=\int_{0}^{1} \dr z' \,\int_0^{+\infty}\dr \epsilon_L \, \delta(\omega_J'-\omega_J-\epsilon_L)\, \mathcal{C}_{i\rightarrow 1}\Big(\{\underline{n}\},z', \omega_J'= \frac{z'\omega_J}{z}, \mu,\mu_\cs\Big) {\cal S}_{1} (n , \epsilon_L,\mu_\cs)  \,, 
\end{align}
with
\bea
\mathcal{S}_1(\epsilon_L) = \text{Tr}\Big[U(n) U(\bar n) \rho_M U(\bar n)^{\dagger}U^{\dagger}(n)\mathcal{M}'\Big].
\label{eq:S1}
\eea
The cs Wilson lines that encode the interaction of the fast-moving color charge partons in the jet with the medium, reads 
\begin{equation}
U(n)\equiv \mathcal{P} \exp\left[ig \int_0^{\infty}  ds  \, n\cdot A_{\rm cs}(s n)\right]\,
\label{eq:CSWilson}
\end{equation}
where $A_{\cs}$ is collinear soft gauge field. Moreover, the function ${\cal S}_1$ now evolves with a SCET Hamiltonian that involves both cs and soft Hamiltonians along with their interaction terms
\begin{align}
\label{eq:H2}
\int {\rm d}t\, H(t) &= \int {\rm d}t \left(H_{\cs}(t)+H_{\s}(t)+ H_{\cs-\s}(t)\right) + \int {\rm d} s\, {\bf O}_{\cs-\s}(sn) \,.
\end{align}
Here, $H_{\cs}$ which describes the dynamics of collinear soft (cs) modes is the standard collinear SCET Hamiltonian and $H_{\s}$ describes the dynamics of the soft partons of the medium, which is the same as the full QCD Hamiltonian. $H_{\cs-\s}$ describes the forward scattering of the collinear-soft gluon in the jet off a soft medium parton and the medium-induced gluon emission is described by the last term as an  operator on the world-line of the hard parton. This operator captures medium-induced radiation to all orders in perturbation theory. 
Therefore, in terms of interaction operators derived in Ref.~\cite{Rothstein:2016bsq}, the operator $\mathcal{O}_{cs-s}$ acquires the form
\begin{align}\label{eq:Oc-s}
{\bf O}_{\cs-\s}(sn) = \int {\rm d}^2\bfq \frac{1}{\bfq^2}\Big[\mathcal{O}^{ba}_{\cs}\frac{1}{\mathcal{P}_{\perp}^2}\mathcal{O}_{\s}^a\Big](sn,\bfq) t^b\,,
\end{align} 
where $\mathcal{P}_{\perp}$ pulls out Glauber momentum from soft operators. Further, $\mathcal{O}_{\cs}$ and $\mathcal{O}_{\s}$ are gauge invariant operators built out of collinear-soft and soft fields, respectively. The $\mathcal{O}_{\cs}$ operator captures the emitted cs gluons generated through Glauber mediated interactions and is given as  
\begin{align}
&\mathcal{O}_{\cs}^{ba} = \frac{8\pi \alpha_s}{\mathcal{P}_{\perp}^2}\Bigg[\mathcal{P}_{\perp}^{\mu}S_n^TW_{n}\mathcal{P}_{\perp \mu} -\mathcal{P}_{\mu}^{\perp}g \mathcal{\tilde B}_{\s \perp}^{n\mu}S_n^TW_n -\nonumber \\
& S_n^TW_ng \mathcal{\tilde B}^{n \mu}_{\perp}\mathcal{P}_{\mu}^{\perp}- g \mathcal{\tilde B}_{\s \perp}^{n\mu}S_n^TW_ng\mathcal{\tilde B}^{n}_{\perp \mu}-\frac{n_{\mu}\bar n_{\nu}}{2}S_n^T ig \tilde G^{\mu\nu}W_n\Bigg]^{ba},
\label{eq:LPT}
\end{align}
\normalsize
where the Wilson line $S_n$ contains cs gluon fields similar to the one in Eq.~\ref{eq:CSWilson} and $S_n^T$ is its conjugate and $W_n$ is the collinear Wilson line.  The interaction of the cs mode with the soft field is given by 
\begin{align}
\label{EFTOp}
{H}_{\cs-\s} 
 &= C_G\frac{i}{2}f^{bcd}\mathcal{B}_{n \perp\mu}^c\frac{\bar n}{2}\cdot(\mathcal{P}+\mathcal{P}^{\dagger})\mathcal{B}_{n \perp}^{d\mu}\frac{1}{\mathcal{P}_{\perp}^2}\mathcal{O}_{\s}^b, 
\end{align}
where the Glauber Wilson coefficient $C_G(\mu)=8\pi\alpha_s(\mu)$ and soft current operator for quarks and gluons $\mathcal{O}_s^b = \sum_{j \in \{q,\bar q, g\}}\mathcal{O}_s^{jb}$ are given as

\small
\begin{equation}
\mathcal{O}_{\s}^{qb}= \bar{\chi}_{s}t^b\frac{\slashed{n}}{2}\chi_{\s}, 
 \   \   \    
\mathcal{O}_{\s}^{gb}=  \frac{i}{2}f^{bcd}\mathcal{B}_{\s \perp\mu}^c\frac{n}{2}\cdot(\mathcal{P}+\mathcal{P}^{\dagger})\mathcal{B}_{\s \perp}^{d\mu} 
\end{equation}
 \normalsize
where the soft operators are dressed with soft Wilson lines. These operators are built out of the gauge invariant building blocks and are defined as
\small
\begin{align}
  \   \   \   \    
  W_{n} &= \text{FT} \  {\bf P} \exp \Big\{ig\int_{-\infty}^0 {\rm d}s \, \bar{n}\cdot A_{n}(x+\bar{n}s)\Big\}  ,
  \nonumber\\
   \   \   \   \    
   S_{n} &= \text{FT} \ {\bf P} \exp \Big\{ig\int_{-\infty}^0 {\rm d}s\, n\cdot A_{\s}(x+s n)\Big\}
   , \nonumber\\
 & \mathcal{B}_{n \perp}^{c\mu} t^c = \frac{1}{g}\Big[W_{n}^{\dagger}iD_{n \perp}^{\mu}W_{ n}\Big],    \  \  
 \mathcal{B}_{\s\perp}^{c\mu} t^c = \frac{1}{g}\Big[S_{n}^{\dagger}iD_{\s \perp}^{\mu} S_{n}\Big] 
 .
\end{align}
\normalsize
Here, FT stands for Fourier transform. Moreover, all the operators encode bare quarks and gluons dressed by Wilson lines.


Note that at this stage, the function ${\cal S}_1$ depends on both the properties of the jet and the medium through the collinear soft and soft modes in it. The two modes cannot be decoupled to all orders in a simple manner. Instead, the factorization of the universal physics associated with the medium from the jet requires us to expand out the operator ${\cal S}_1$ order-by-order in the number of Glauber interactions between the jet and the medium. We can therefore write the single subjet function as a series in the number of Glauber operators 
\bea 
{\cal S}_{1} = \sum_{i=0}^{\infty}{\cal S}_{1}^{(i)}
\label{eq:S1Series}
\eea 
In this series, ${\cal S}_{1}^{(0)}$ is the vacuum contribution, which was thoroughly discussed in Ref.~\cite{Mehtar-Tani:2024smp} and contributes to the resummation of threshold logarithms $\ln (1-z)$. Moreover, ${\cal S}_{1}^{(1)}$ is the single scattering regime with leading order Glauber interaction contributions. 
In this paper, we focus on the terms with arbitrary number of interactions of the jet with the medium. 
For the term with $n$ interactions at  ${\cal O}(n)$ with $n>0$, we can write 
\begin{align}\label{eq:FctAl}
&{\cal S}_{1}^{(n)}(\epsilon_L, \mu) = |C_{G}|^{2n}\Bigg[\prod_{i=1}^{n}\int_{0}^L {\rm d}x^-_i\Theta(x^-_i-x_{i+1}^-)\int \frac{{\rm d}^2\bfk_{i}}{(2\pi)^3}\nonumber \\
&\varphi(\bfk_{i},\mu,\nu,x^-_i)\Bigg]
{\bf F}_{1}^{(n)}(\epsilon_L; \bfk_{1}, \ldots, \bfk_{n}; x_1^-, \ldots x_n^-;\nu)\,.
\end{align}
Note that this expression contains $n$ copies of the same two point medium correlator $\varphi$, which is defined as
\begin{align}
&\varphi(\bfk,\mu,\nu)= \frac{1}{\bfk^2}\frac{1}{N_c^2-1}\int \frac{{\rm d}k^-}{2\pi}\int {\rm d}^4r\, e^{-i \bfk \cdot \bfr+ik^-r^+}\text{Tr}\Big[e^{-i\int {\rm d}t H_s(t) }\mathcal{O}^A_{\s}(r) \rho_M \mathcal{O}^A_{\s}(0)e^{i\int {\rm d}t H_s(t) }\Big],\,\qquad
\label{eq:medcorr}
\end{align}
where $\rho_M$ is thermal/medium density matrix.  The result in Eq.~\ref{eq:FctAl} is valid when the mean free path of the jet $\lambda_{\rm mfp}$ is much larger than the color screening length $1/m_D$ in the medium. Since the mean free path is an emergent scale, we will verify this assumption through an explicit computation. Below we discuss both dilute and dense medium with multiple scattering cases in more details.


\section{Dilute medium}
\label{sec:DilM}
For the case of a dilute medium, we can truncate the series in Eq.~\ref{eq:S1Series} to just the first two terms, corresponding to vacuum and single interaction with the medium. As mentioned in the previous section, this is justified when the mean free path of the jet, which in part is controlled by the medium density, is much larger than the medium size $L$. Although our ultimate goal in this paper is the analysis of the dense medium, the dilute limit will be valuable in understanding the higher order radiative corrections that give rise to large logarithms in a single interaction. These radiative corrections need to be resummed to maintain the accuracy of the calculation and will also be present for the case of a dense medium.  
\subsection{Single subjet single interaction}

We have already computed the single subjet for the case of zero and single interaction.  The case of zero interaction is just the vacuum result and gives the threshold logarithms and was presented in \cite{Mehtar-Tani:2024smp}.  The single interaction with the medium leads to the soft limit of the GLV which has both the Lipatov and broadening contributions. 

Here we give here the expressions for the medium induced collinear soft function for a single interaction with the medium i.e, the function ${\bf F}_{1}^{(1)}$. 

\begin{figure}
\centering
\includegraphics[width=0.7\linewidth]{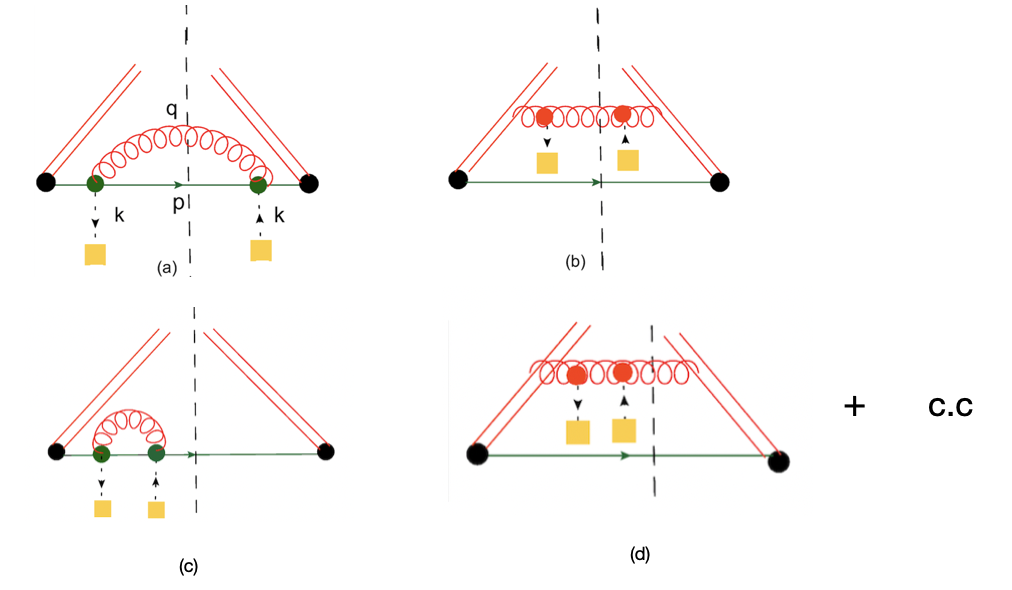}
\caption{Feynman diagrams for single interaction and single gluon emission~\label{fig:1int}. The red parallel lines indicate the $U(n)$ collinear soft Wilson line. The red gluon emission is a collinear soft gluon. The green vertex is the Glauber Lipatov vertex which encodes medium induced radiation, the red vertex encodes forward scattering of the gluon  with the medium while the golden square indicates a soft scattering center in the medium.}
\end{figure}

In this paper, we will only consider the regime when the formation time of the gluon  emitted with energy $E$ and a transverse momentum $\bfq$ given by $t_f \sim E/\bfq^2$ is much smaller than the medium size $L$. This is equivalent to taking the $ L \rightarrow \infty$ limit while evaluating all relevant Feynman diagrams. At the end of this paper, we will revisit the validity of this assumption and discuss the regime where it breaks down. In this case the diagrams that contribute to the final result are drastically reduced. The relevant diagrams that  contribute are shown in Fig.\ref{fig:1int}
The result for the single interaction can be written as 
\begin{align}\label{eq:F1q}
&{\bf F}_{1,q}^{(1)}(\epsilon_L,{\bfk},\mu)=\frac{\alpha_s(N_c^2-1)}{4\pi^2}\mu^{2\epsilon} \int \frac{{\rm d}^{2-2\epsilon}\bfq}{\bfk^2 \bfq^2} \frac{2 {\bfk} \cdot {\bfq}}{({\bfq}+{\bfk})^2}  \int\frac{{\rm d}q^-}{q^-}\Theta\left(|\bfq|- \frac{q^-R}{2}\right)\Big[\delta(\epsilon_L)-\delta(q^--\epsilon_L)\Big].
\end{align}
This result agrees exactly with the one loop result in the soft limit in the GLV formalism ~\cite{Gyulassy:2000fs,Gyulassy:2000er,Wiedemann:2000za} which is reasonably good cross check on our framework. 

We note some important points about this result: The scaling of $q^-$ is not a kinematic constant but is dynamically determined through transverse momentum $\bfq$ transferred to the cs parton by the medium. In a single interaction this is set by the distribution of $|\bfk|$ which peaks at $|\bfk| \sim m_D$. Hence for dilute medium where single interaction is sufficient, $ \bfq \sim m_D$ and $q^- \sim m_D/R$. Moreover, the radiative corrections to the leading order result involves the full Balitsky-Fadin-Kuraev-Lipatov (BFKL) evolution. Since $\bfq \sim m_D$ is a non-perturbative scale, the function ${\cal S}_{1}^{(1)}(\epsilon_L,R)$  is fully non-perturbative and depends on the radius of the jet R but is independent of the jet energy. 

\subsection{Resummation}
For a dilute medium, the function ${\cal S}_{1}^{(1)}(\epsilon_L,R)$ is fully non-perturbative and a perturbative BFKL resummation is therefore not valid. However, we will see that in a dense medium, with multiple interactions there is possibility of raising the virtuality of the collinear soft radiation to a perturbative scale where a perturbative analysis will be needed. In anticipation of this result, we present here the leading order BFKL resummation for the  single interaction case. 
We present the final result for the BFKL resummed collinear soft function. Details can be found in Ref.~\cite{Singh:2024vwb}
\begin{equation}
{\bf F}^{(1)}_{1,{\rm R}}(\bfk,\mu,\nu_f)=\sum_{n=-\infty}^{\infty} \int \dr^2l_{\perp}{\bf F}^{(1)}_1(l_{\perp},\mu,\nu_0)\int\frac{\dr\nu}{2\pi}k_{\perp}^{-1+2i\nu}l_{\perp}^{-1-2i\nu}e^{in(\phi_k-\phi_l)}e^{-\frac{\alpha_s(\mu)N_c}{\pi}\chi(n,r)\log\frac{\nu_f}{\nu_0}}.  
\label{eq:resumS}
\end{equation}
Here $\nu_f$ is the natural rapidity scale for the collinear soft function ${\bf F}_1^{(1)}$ which is  $\epsilon_L $, while $\nu_0 \sim k_{\perp}^2/m_D \sim m_D$. Just as in the case of DIS at small $x$, this BFKL evolution resums the $\ln 1/x$, where 
\bea
x = \frac{\text{probe virtuality}^2}{\text{c.o.m. energy}^2} \sim \frac{k_{\perp}^2}{m_D \epsilon_L} \sim \frac{m_D}{\epsilon_L}.
\eea
We can therefore write the resummed single subjet function as
\begin{align}
{\cal S}_1^{(1)} &= L \int \frac{\dr^2k_{\perp}}{(2\pi)^3}\varphi(k_{\perp},\nu=\nu_0){\bf F}^{(1)}_{1,{\rm R}}(k_{\perp},\mu,\nu_f) \nonumber \\
&= L\int \dr^2l_{\perp}\sum_{n=-\infty}^{\infty}  \int \frac{\dr^2\bfk}{(2\pi)^3}\varphi(\bfk,\nu=\nu_0){\bf F}^{(1)}_1(l_{\perp},\mu,\nu_0)\int\frac{d\nu}{2\pi}k_{\perp}^{-1+2i\nu}l_{\perp}^{-1-2i\nu}\nonumber\\
&\qquad e^{in(\phi_k-\phi_l)}e^{-\frac{\alpha_s(\mu)N_c}{\pi}\chi(n,r)\log\frac{\nu_f}{\nu_0}}.  
\end{align}
For later convenience, we interchange the variables $k_{\perp} \longleftrightarrow l_{\perp}$, $\nu \rightarrow -\nu$ and $n \rightarrow -n$ to absorb the resummed factor inside the medium correlator $\varphi$ to write
\bea
{\cal S}_1^{(1)}(\mu \sim \epsilon_L R)  &= &L\int \frac{d^2\bfk }{(2\pi)^3}\varphi_{R}(k_{\perp},\nu \sim \nu_{cs}){\bf F}^{(1)}_1(k_{\perp},\mu \sim \epsilon_L R,\nu_{cs}; \epsilon_L, R),
\eea
where the collinear soft function is now evaluated with all of its rapidity logs minimized. Note that this function also has threshold logarithms which only appear at two loops and can also be resummed systematically by running in $\mu$. We choose to run the hard collinear function $\mathcal{C}_{i \rightarrow 1}$ form the scale $\mu \sim p_TR$ down to the scale $\epsilon_L R$ to minimize all threshold logarithms in  ${\bf F}^{(1)}_1$. With these choices, hereafter we can evaluate  ${\bf F}^{(1)}_1$ at leading order, i.e. Eq.\ref{eq:F1q} .


\section{Dense medium}
\label{sec:DenseM}
\subsection{Single subjet multiple interactions}

In this section we consider the case of multiple interactions of a collinear-soft gluon with the medium. The collinear-soft gluon can either be emitted through  vacuum evolution from Wilson lines or through medium induced radiation. Once emitted at either source, this gluon then undergoes multiple scatterings in the medium undergoing broadening. We will also consider subsequent radiative corrections that are induced by anomalous dimensions of the factorized functions. We will focus on the large but dense medium regime in the Markovian limit so that quantum interference between successive interactions, namely the LPM effect can be ignored. This is valid whenever the formation time of the radiated gluons $t_f$ is much smaller than the medium size. The scale $t_f \sim \epsilon_L/\bfq_i^2$ is decided by its initial transverse momentum $\bfq_i$ and energy $\epsilon_L$  at the \textit{source} of the emission. On the other hand, the only constraint on the phase space that contributes to the measurement is that the ratio of the \textit{final} transverse momentum of the gluon $\bfq_{f}$ to its energy should scale as the jet radius R, i.e. $\bfq_{f}/\epsilon_L \sim R$. As has been noted earlier in literature~\cite{Casalderrey-Solana:2011ule}, multiple interactions of a gluon in the medium can lead to an emergent transverse momentum scale $Q_{\med}$ which can be much larger than the medium scale $m_D$ which can lead to small formation times. In this paper we therefore  start off with the assumption that the lifetime of the gluon radiation is small and then check through an explicit calculation the self consistency of this assumption. 

This is a simpler case which will help us establish the nature of the emergent scale. At the end of the paper, we will discuss the regime where the Markovian limit breaks down and leave the computation with full quantum interference effects for the future. 
In the Markovian approximation, we can consider the vacuum induced and medium induced cs radiation independently since there are no interference terms between them.
\subsubsection{Broadening of vacuum induced radiation}
We first compute the broadening of the cs gluon sourced by the vacuum, i.e., the collinear soft Wilson line. 
For single interaction with the medium, the broadening term is given as 
\begin{align}
&{\bf F}_{B}^{(1)}(\epsilon_L,{\bfk})=\frac{\alpha_s(N_c^2-1)}{4\pi^2}\mu^{2\epsilon} \int \frac{{\rm d}^{d}\bfq}{\bfk^2 }\Big[ \frac{1}{({\bfq}+{\bfk})^2} - \frac{1}{{\bfq}^2}\Big]\int\frac{{\rm d}q^-}{q^-}\Theta\left(|\bfq|- \frac{q^-R}{2}\right)\!\Big[\delta(\epsilon_L)-\delta(q^--\epsilon_L)\Big],
\end{align}
where the dimension $d=2-2\epsilon$. In order for eventually resum the arbitrary number of jet medium interactions, we rewrite the above equation in the impact parameter space 
\begin{align}
{\bf F}_{B}^{(1)}(\epsilon_L,{\bfk})&=\frac{\alpha_s(N_c^2-1)}{4\pi^2}\mu^{2\epsilon} \int \frac{{\rm d}^{d}\bfq}{\bfk^2 }\int\frac{{\rm d}q^-}{q^-}\Theta\left(|\bfq|- \frac{q^-R}{2}\right)\Big[\delta(\epsilon_L)-\delta(q^--\epsilon_L)\Big]\nonumber \\
&\qquad\quad\, \int \frac{d^2\bfp}{(2\pi)^2}\int \frac{d^2 \bfb e^{i \bfp \cdot \bfb}}{(\bfp+ \bfq)^2}\left(e^{-i \bfk \cdot \bfb}-1\right).
\end{align}
\begin{figure}
\centering
\includegraphics[width=\linewidth]{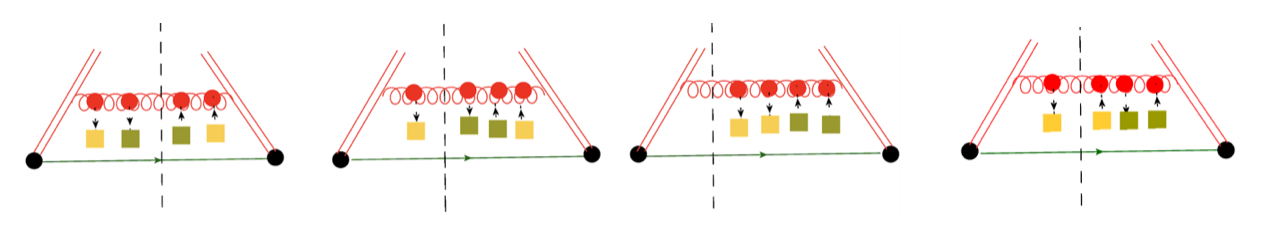}
\caption{Feynman diagrams for two scatterings of a vacuum induced collinear soft gluon emission~\label{TISE}.}
\end{figure}
The diagrams contributing to multiple scattering amplitudes of cs radiation off medium partons are shown in Fig. \ref{TISE}.   
In the $L \rightarrow \infty$ limit, the cs parton goes on-shell between successive interactions. Using the Feynman rules the amplitude for two successive scatterings reads as
\begin{align}
{\bf F}_{B}^{(2)}(\epsilon_L,\bfk_1, \bfk_2)&=\frac{\alpha_s(N_c^2-1)}{4\pi^2}\mu^{2\epsilon} \int \frac{{\rm d}^{d}\bfq}{\bfk_1^2 \bfk_2^2}\int\frac{{\rm d}q^-}{q^-}\Theta\left(|\bfq|- \frac{q^-R}{2}\right)\Big[\delta(\epsilon_L)-\delta(q^--\epsilon_L)\Big]\nonumber \\
&\quad\, \int \frac{d^2\bfp}{(2\pi)^2}\int \frac{d^2 \bfb e^{i \bfp \cdot \bfb}}{(\bfp+ \bfq)^2}\left(e^{-i \bfk_1 \cdot \bfb}-1\right)\left(e^{-i \bfk_2 \cdot \bfb}-1\right)
\end{align}
where $\bfb$ is impact parameter.  $\bfk_1$ and $\bfk_2$ are exchanged Glauber momentum. We can generalize this expression for an arbitrary number of Glauber gluon exchanges between the jet and the medium that reads as
\begin{align}
{\bf F}_{B}^{(n)}(\epsilon_L,\bfk_1, \bfk_2,...,\bfk_n)&=\frac{\alpha_s(N_c^2-1)}{4\pi^2}\mu^{2\epsilon} \int {\rm d}^{d}\bfq\int\frac{{\rm d}q^-}{q^-}\Theta\left(|\bfq|- \frac{q^-R}{2}\right)\Big[\delta(\epsilon_L)-\delta(q^--\epsilon_L)\Big]\nonumber \\
&\quad\, \int \frac{d^2\bfp}{(2\pi)^2}\int \frac{d^2 \bfb e^{i \bfp \cdot \bfb}}{(\bfp+ \bfq)^2}\prod_{i=1}^n\frac{\left(e^{-i \bfk_i \cdot \bfb}-1\right)}{\bfk_i^2}
\end{align}
where the product is over the number of exchanged Glauber gluons. We can sum over all the arbitrary interactions contributing to the broadening of vacuum cs radiation and obtain the one loop subjet function  while also including BFKL resummation. The result,  excluding the vacuum contribution ${\cal S}_1^{(0)}$ can be written in a compact form as an exponential.
\begin{align}
{\cal S}_{1,B}( \mu\sim \epsilon_L R) &= \frac{\alpha_s(N_c^2-1)}{4\pi^2}\mu^{2\epsilon} \int {\rm d}^{d}\bfq\int\frac{{\rm d}q^-}{q^-}\Theta\left(|\bfq|- \frac{q^-R}{2}\right)\Big[\delta(\epsilon_L)-\delta(q^--\epsilon_L)\Big]\nonumber \\
& \int \frac{d^2\bfp}{(2\pi)^2}\int \frac{d^2 \bfb\,e^{i \bfp \cdot \bfb}}{(\bfp+ \bfq)^2}\Big[\exp{\Big\{-L\int \frac{d^2\bfk}{\bfk^2}\left(1-e^{-i \bfk \cdot \bfb}\right)\varphi_R(\bfk, \nu \sim \nu_{\cs}) \Big\}}-1\Big]
\label{eq:S1B}    
\end{align}
where $\varphi_R$ is the resummed medium correlator defined in Eq.~\ref{eq:medcorr} and $\nu_{\cs} \sim \epsilon_L$ is rapidity scale for single subjet function. 
\subsubsection{Broadening for medium induced radiation}

\begin{figure}
\centering
\includegraphics[width=0.5\linewidth]{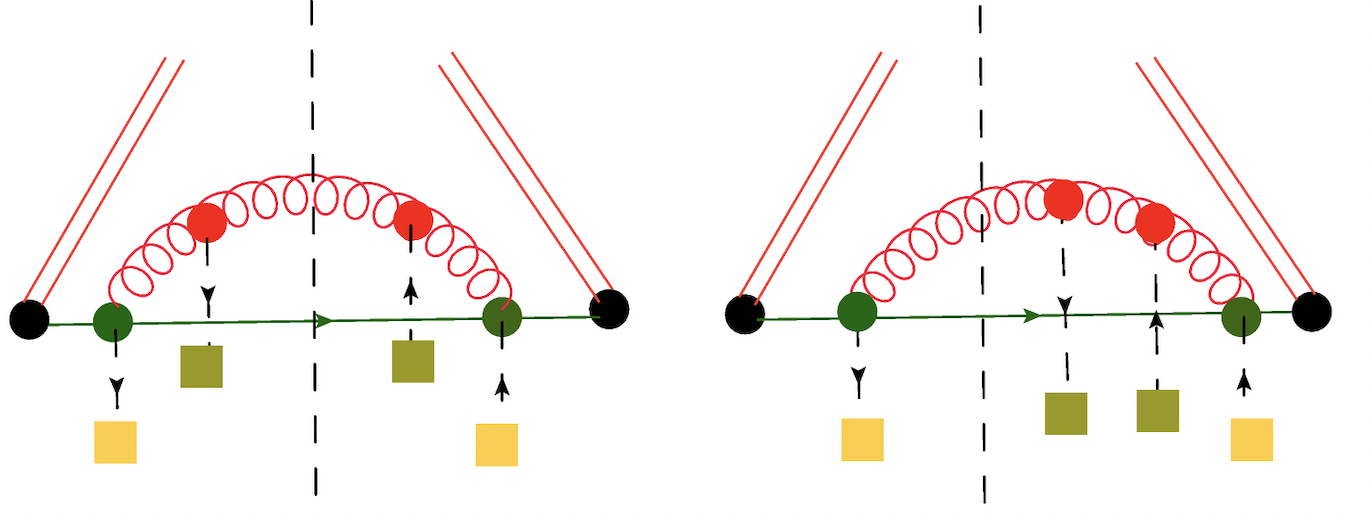}
\caption{Feynman diagrams for medium scattering of a medium induced collinear soft gluon ~\label{MIDI}.}
\end{figure}

Next we consider contribution to the broadening from medium induced radiation. In this case first the cs gluon is emitted through the interaction of collinear mode with the medium and subsequently undergoes multiple scatterings with the medium soft partons. Therefore, the leading order term in the jet function describes the production of cs gluon through the Lipatov vertex and is given as 
\begin{align}
&{\bf F}_{M}^{(1)}(\epsilon_L,{\bfk})=-\frac{\alpha_s(N_c^2-1)}{4\pi^2}\mu^{2\epsilon} \int \frac{{\rm d}^{d}\bfq}{\bfq^2 (\bfk+\bfq)^2}\int\frac{{\rm d}q^-}{q^-}\Theta\left(|\bfq|- \frac{q^-R}{2}\right)\Big[\delta(\epsilon_L)-\delta(q^--\epsilon_L)\Big]. 
\end{align}
In the number of interactions with the medium, the next-to-leading (NLO) diagrams are shown in Fig.\ref{MIDI}. The result for two Glauber exchanges can then be written as 
\begin{align}
{\bf F}_{M}^{(2)}(\epsilon_L,\bfk_1, \bfk_2)&=-\frac{\alpha_s(N_c^2-1)}{4\pi^2}\mu^{2\epsilon} \int \frac{{\rm d}^{d}\bfq} {\bfq^2 (\bfk_1+\bfq)^2 \bfk_2^2}\int\frac{{\rm d}q^-}{q^-}\Big[\delta(\epsilon_L)-\delta(q^--\epsilon_L)\Big]\nonumber \\
& \Bigg[ \Theta\left(|\bfq +\bfk_2|- \frac{q^-R}{2}\right)-\Theta\left(|\bfq|- \frac{q^-R}{2}\right) \Bigg].
\end{align}
Similar to the previous case, in order to resum arbitrary number of Glauber insertions from the medium we can further simplify the above equation by writing this in impact parameter space
\begin{align}
{\bf F}_{M}^{(2)}(\epsilon_L,\bfk_1, \bfk_2)&=-\frac{\alpha_s(N_c^2-1)}{4\pi^2}\mu^{2\epsilon} \int \frac{{\rm d}^{d}\bfq }{\bfq^2 (\bfk_1+\bfq)^2 \bfk_2^2}\int\frac{{\rm d}q^-}{q^-}\Big[\delta(\epsilon_L)-\delta(q^--\epsilon_L)\Big]\nonumber \\
&  \int \frac{\dr^2\bfp}{(2\pi)^2}\int \dr^2 \bfb\, e^{i \bfp \cdot \bfb}\,\Theta\left(|\bfq +\bfp|- \frac{q^-R}{2}\right)\left(e^{-i \bfk_2 \cdot \bfb} -1 \right).
\end{align}
Generalizing the above equation for $n$ number of Glauber insertions by adding higher order diagrams, always working in the Markovian limit, we obtain
\begin{align}
{\bf F}_{M}^{(n)}(\epsilon_L,\bfk_1, \bfk_2, .. \bfk_n)&=-\frac{\alpha_s(N_c^2-1)}{4\pi^2}\mu^{2\epsilon} \int \frac{{\rm d}^{d}\bfq }{\bfq^2 (\bfk_1+\bfq)^2 }\int\frac{{\rm d}q^-}{q^-}\Big[\delta(\epsilon_L)-\delta(q^--\epsilon_L)\Big]\nonumber \\
&  \int \frac{\dr^2\bfp}{(2\pi)^2}\int \dr^2 \bfb\, e^{i \bfp \cdot \bfb}\,\Theta\left(|\bfq +\bfp|- \frac{q^-R}{2}\right)\prod_{i=2}^n\frac{\left(e^{-i \bfk_i \cdot \bfb} -1 \right)}{{\bfk_i^2}}.
\end{align}
Finally, combining the above equation with the medium correlator defined in Eq.~\ref{eq:medcorr}, the all order broadening term contributing to single subjet function from the medium-induced radiation can be written as
\begin{align}
&{\cal S}_{1,M}(\epsilon_L, \mu)= -\frac{\alpha_s(N_c^2-1)\mu^{2\epsilon}}{4\pi^2}\int_0^L\! \dr x^- \! \int\! \frac{\dr^2 \bfu}{\bfu^2} \varphi_R(\bfu, \nu_{\cs})\! \int \frac{\bfu^2 \,{\rm d}^{d}\bfq}{\bfq^2 (\bfu+\bfq)^2 }\!\int\frac{{\rm d}q^-}{q^-}\Big[\delta(\epsilon_L)-\delta(q^--\epsilon_L)\Big]\nonumber\\
&\quad \int \frac{\dr^2\bfp}{(2\pi)^2}\int \dr^2 \bfb e^{i \bfp \cdot \bfb}\Theta\left(|\bfq +\bfp|- \frac{q^-R}{2}\right)\left(1+ \sum_{n= 1}^{\infty}\prod_{i=1}^n \frac{(x^-)^n}{n!}\int  \frac{\dr^2\bfk_i}{\bfk_i^2}\left(e^{-i \bfk_i \cdot \bfb} -1 \right)\varphi_R(\bfk_i, \nu_{\cs})\right)    
\end{align}
where same as previous case the rapidity scale in the medium correlator is $\nu\sim\nu_{\cs} \sim \epsilon_L$ and the scale $\mu\sim \epsilon_L R$.  The above equation can now be exponentiated to obtain
\begin{align}
&{\cal S}_{1,M}(\epsilon_L, \mu)= -\frac{\alpha_s(N_c^2-1)\mu^{2\epsilon}}{4\pi^2}\int_0^L \dr x^-  \int \frac{\dr^2 \bfu}{\bfu^2} \varphi(\bfu, \nu_{\cs}) \int \frac{\bfu^2\,{\rm d}^{d}\bfq }{\bfq^2 (\bfu+\bfq)^2 }\int\frac{{\rm d}q^-}{q^-}\Big[\delta(\epsilon_L)-\delta(q^--\epsilon_L)\Big]\nonumber \\
&\quad\int \frac{\dr^2\bfp}{(2\pi)^2}\int \dr^2 \bfb\, e^{i \bfp \cdot \bfb}\,\Theta\left(|\bfq +\bfp|- \frac{q^-R}{2}\right)\exp{\Bigg\{-x^-\int  \frac{\dr^2\bfk}{\bfk^2}\left(1-e^{-i \bfk \cdot \bfb} \right)\varphi_R(\bfk, \nu_{\cs} )\Bigg\}}.   
\end{align}
We can now explicitly perform the integration over $x^-$ to write broadening of medium-induced emission as  
\bea
&&{\cal S}_{1,M}(\epsilon_L, \mu )= \frac{\alpha_s(N_c^2-1)}{4\pi^2}\mu^{2\epsilon} \int \frac{\dr^2 \bfu}{\bfu^2} \varphi_R(\bfu, \nu_{cs}) \int \frac{\bfu^2\,{\rm d}^{d}\bfq  }{\bfq^2 (\bfu+\bfq)^2 }\int\frac{{\rm d}q^-}{q^-}\Big[\delta(\epsilon_L)-\delta(q^--\epsilon_L)\Big]\nonumber \\
&&\int \frac{\dr^2\bfp}{(2\pi)^2}\int \dr^2 \bfb\, e^{i \bfp \cdot \bfb}\,\Theta\left(|\bfq +\bfp|- \frac{q^-R}{2}\right)\frac{1- \exp{\Big\{-L\int  \frac{\dr^2\bfk}{\bfk^2}\left(1-e^{-i \bfk \cdot \bfb} \right)\varphi_R(\bfk, \nu_{\cs} )\Big\}}}{\int\frac{\dr^2\bfl}{\bfl^2}\left(e^{-i \bfl \cdot \bfb} -1\right)\varphi_R(\bfl, \nu_{\cs})}.
\label{eq:S1M}
\eea
\section{An estimate of the emergent scale}
\label{sec:eScale}
\begin{figure}[h!]
\centering
\includegraphics[scale=0.41]{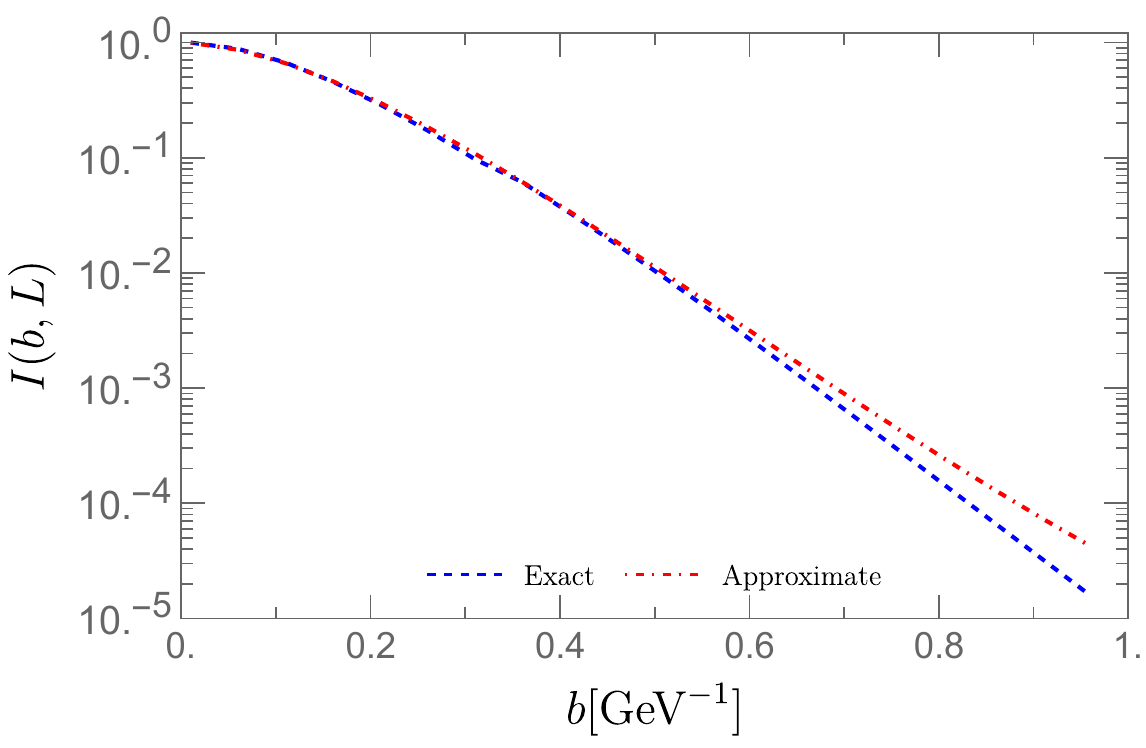}
\includegraphics[scale=0.41]{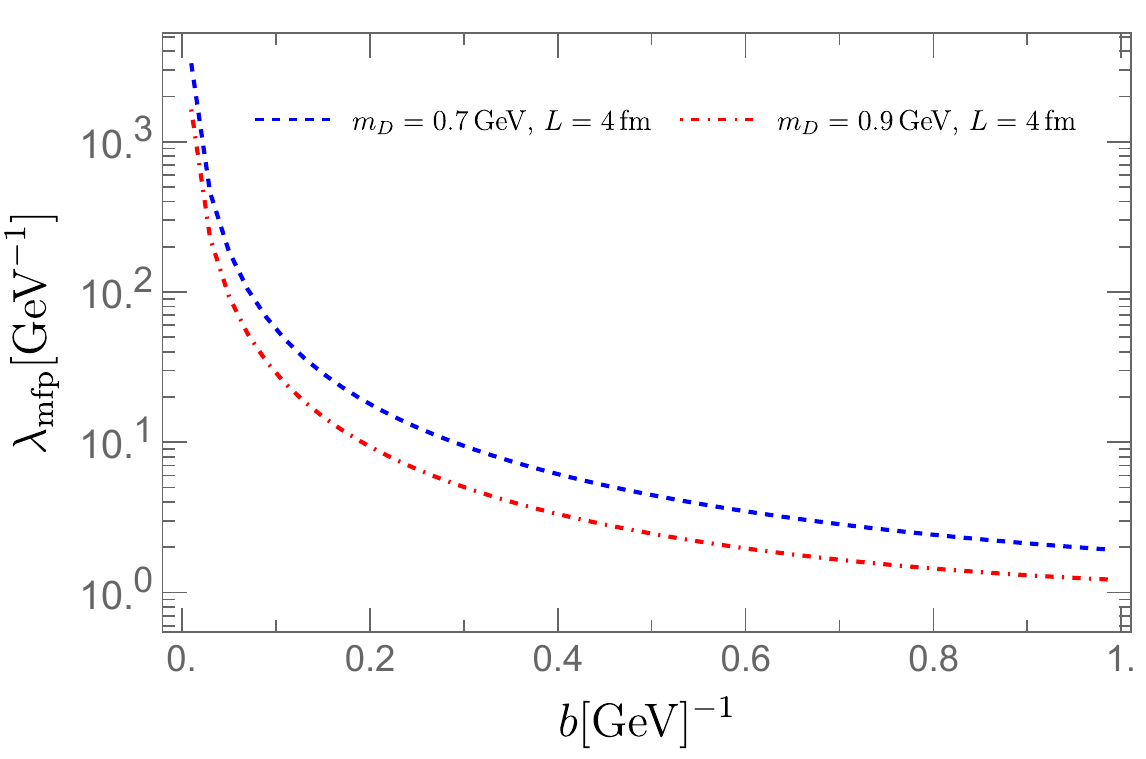}
\caption{(a) Comparison of the exact (Eq.~\ref{eq:Ib}) and approximate functions (Eq.~\ref{eq:IbA}) $I(b,L)$ as a function of impact parameter $b$ for $L=4$ fm and $m_D=0.7$ GeV. (b) The mean free path, defined in Eq.~\ref{eq:mfp}, as a function of impact parameter for $L=4$ fm and two different value of Debye mass.}
\label{fig:Ib}
\end{figure}

We can now obtain the full collinear soft function in the Markovian limit by adding the two contributions from broadening of vacuum induced and medium induced radiation. With some simplifications this takes the form
\begin{align}
&{\cal S}_{1}(\epsilon_L, \mu)={\cal S}_{1,B}(\epsilon_L, \mu )+{\cal S}_{1,M}(\epsilon_L, \mu) \nonumber \\
&\quad= \frac{2\alpha_s}{\pi^2}\mu^{2\epsilon}  \int\frac{{\rm d}q^-}{q^-}\Big[\delta(\epsilon_L)-\delta(q^--\epsilon_L)\Big]\int \frac{\dr^2\bfp}{(2\pi)^2}\int \dr^2 \bfb\, e^{i \bfp \cdot \bfb}\int \frac{{\rm d}^{d}\bfq  }{\bfq^2 }\Theta\left(|\bfq +\bfp|- \frac{q^-R}{2}\right)\nonumber \\
&\qquad\,\quad\Bigg[1 -\frac{\int \frac{\dr^2 \bfu}{(\bfu+\bfq)^2} \varphi_R(\bfu, \nu_{\cs} )}{\int\frac{\dr^2\bfl}{\bfl^2}\left(e^{-i \bfl \cdot \bfb} -1\right)\varphi_R(\bfl, \nu_{\cs})}\Bigg] \Bigg[\exp{\Big\{-L\int  \frac{\dr^2\bfk}{\bfk^2  }\left(1-e^{-i \bfk \cdot \bfb} \right)\varphi_R(\bfk, \nu_{\cs})\Big\}}-1 \Bigg] 
\label{eq:S1B&M}
\end{align}
Let us note that due to multiple scatterings,   even with the summation over arbitrary number of scatterings the scaling for  absolute value of $\bfq$ and hence  $q^-$ is still undetermined; only the scaling for their ratio $\bfq/q^- \sim R$ is set by the jet radius. For the case of a single interaction, the scale for $\bfq$ is set by the scale for $\bfk \sim m_D$. However for multiple scatterings this is dictated by the average transverse momentum imparted by the medium to the jet. Intuitively we can guess that in a dense medium, the net transverse momentum gained by the parton will be typically much higher than $m_D$ and will depend on the density of the medium and the effective strength of the interaction of the jet with the medium partons. In literature~\cite{Casalderrey-Solana:2011ule} this is given by the parameter $\sqrt{\hat q L}$. Here we want to see how such an object can be defined in terms of the operators in our EFT framework. 
The factor which determines this scale is the length dependent exponent,
\bea 
I(b,L)= \exp{\Big\{-L\int  \frac{\dr^2\bfk}{\bfk^2  }\left(1-e^{-i \bfk \cdot \bfb} \right)\varphi_R(\bfk, \nu_{\cs} )\Big\}}  \equiv e^{-L M(b)}, 
\eea
where 
\bea
M(b)&=& \int  \frac{\dr^2\bfk}{\bfk^2  }\left(1-e^{-i \bfk \cdot \bfb} \right)\varphi_R(\bfk, \nu_{\cs})=  2\pi \int \frac{k_{\perp} \dr k_{\perp}}{\bfk^2+ m_D^2}\varphi_R(\bfk, \nu_{\cs})\left(1-J_0(b|\bfk|)\right),
\eea
where $k_{\perp}=|\bfk|$. To get a sense of how this object behaves as a function of $b = |\bfb|$, we borrow some intuition from the weak coupling behavior of the function $\varphi(\bfk, \nu_{\cs})$. Through explicit computation~\cite{Singh:2024vwb} in a thermal medium, we know that at tree level to a very good approximation  $\varphi = \varphi_0/\bfk^2$  where $\varphi_0$ is a constant with dimensions $[M]^3$. Since $\varphi_0$ only knows about the scale $T \sim m_D$, as a brute force estimate we can take $\varphi_0 \sim c\, m_D^3$\footnote{A tree level computation yields  $\varphi_0 =g^4T^3$ which for a strongly coupled medium ,i.e. $g \sim 1$ scales as $m_D^3$} where $c$ is a constant which we will ignore in this study.  
\bea 
M(b) \approx \frac{\pi \varphi_0}{m_D^2}(1- m_Db K_1(m_Db)) \approx \pi m_D(1- m_Db K_1(m_Db)), 
\eea
where $K_1(m_Db)$ is the modified Bessel function of the first kind. Therefore 
\bea
I(b,L) \approx  \exp{\Big\{-L\pi m_D(1- m_Db K_1(m_Db)) \Big\}}. 
\label{eq:Ib}
\eea
For typical phenomenological values in current heavy ion collisions, $m_D \sim 0.7$ GeV and $L \sim 4 $ fm. For these values, we observe that the co-efficient in the exponent $L\pi m_D \approx 43$ which is very large suggesting a rapid decay with increasing $b$. Hence we expect that relevant scale for $b$ $\ll 1/m_D$. In this limit 
\bea 
I(b,L) &&\approx \exp{\Bigg\{-b^2\frac{L}{2}\pi m_D^3 \ln \frac{2\sqrt{e}}{m_Dbe^{\gamma_E}}\Bigg\}} +O(m_D^2b^2),\nonumber \\
& \equiv & \exp{\Bigg\{-b^2L \hat q \ln \frac{2\sqrt{e}}{m_Dbe^{\gamma_E}}\Bigg\}}. 
\label{eq:IbA}
\eea
 We see that this approximates the full result almost exactly over the range where it would contribute to the eventual $b$ integral as shown in the left panel of Fig.\ref{fig:Ib} . The prefactor the logarithm is usually referred to as the jet transport or $\hat q^2$ parameter in literature and has the units of transverse momentum squared per unit length. The value we have used here  is just a rough estimate based on dimensional analysis and we will come back to the all orders operator definition later in the paper. 
 The analysis so far  gives a number $\sqrt{\hat q L} \sim 3$ GeV which is a perturbative scale. 
We can also write this function as 
\bea
I(b,L) \equiv \exp{\Bigg\{-\frac{L}{\lambda_{\text{mfp}}(b)} \Bigg\}} 
\label{eq:mfp}
\eea
where we can interpret $\lambda_{\text{mfp}}(b)$ as the mean free path of the cs parton. Due to the nature of this function, we know that only the region of $b \ll 1/m_D \in \{0, 0.5\}$ contributes to the final result. We can then plot the mean free path over this region as shown in the right panel of Fig. \ref{fig:Ib}. We clearly see that $\lambda_{\text{mfp}} \gg 1/m_D$, which supports our assumption that the successive interactions of the jet parton occurs with independent or color uncorrelated  medium soft partons.

\subsection{Analysis for vacuum induced radiation}
\begin{figure}
\centering
\includegraphics[width=0.7\linewidth]{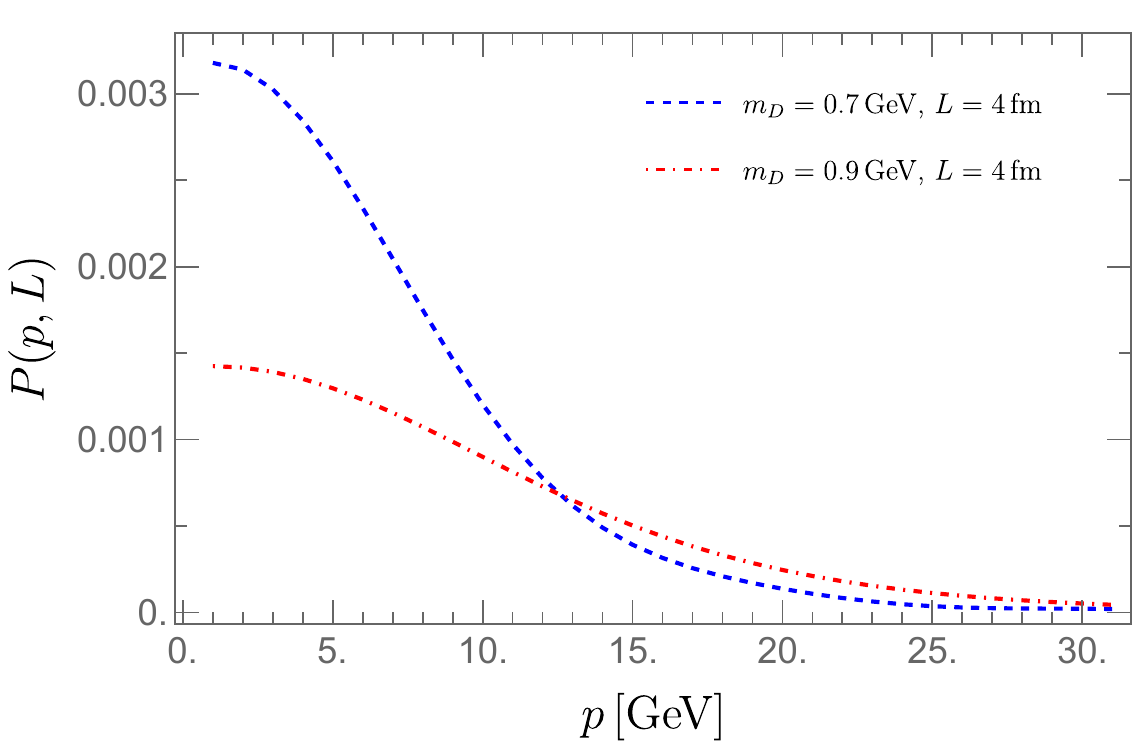}
\caption{The probability distribution defined in Eq.~\ref{eq:ProbB} as a function of transverse momentum $p=|\bfp|$ for $L=4$ fm and two different value of the Debye mass for broadening of a gluon in a dense medium in the Markovian limit.}
\label{fig:Pp}
\end{figure}

In the previous section we have invoked the existence of an intrinsic medium scale $\sqrt{\hat q L}$ conjugate to the impact parameter $b$. 
Next we want to understand how this analysis translates to a scale for $|\bfq| \sim Q_{\text{med}}$, the average transverse momentum gained by a gluon in the medium. To estimate this scale, we first consider the broadening of the cs radiation sourced by vacuum evolution. We will subsequently use this scale to comment on the nature of non-perturbative physics for both vacuum and medium induced radiation. 
The single subjet function ${\cal S}_{1,B}$ with the broadening of vacuum induced radiation to all orders obtained in Eq.~\ref{eq:S1B} along with the approximations discussed in the previous section reads as
\begin{align}
{\cal S}_{1,B}(\epsilon_L, \mu)
& \approx \frac{\alpha_s(N_c^2-1)}{4\pi^2}\mu^{2\epsilon}  \int\frac{{\rm d}q^-}{q^-}\Big[\delta(\epsilon_L)-\delta(q^--\epsilon_L)\Big]\int \frac{\dr^2\bfp}{(2\pi)^2}\int \frac{{\rm d}^{d}\bfq  }{|\bfq+ \bfp|^2 }\Theta\left(|\bfq| - \frac{q^-R}{2}\right)\nonumber \\
& \int \dr^2 \bfb\, e^{i \bfp \cdot \bfb}\Bigg[\exp{\Bigg\{-\frac{\pi}{2}b^2L m_D^3 \ln \frac{2\sqrt{e}}{m_Dbe^{\gamma_E}}\Bigg\}} -1 \Bigg]    
\end{align}
where the scale $\mu\sim  \sim \epsilon_L R$. We first focus on the term with integration over the variable $b$ and perform all integrations to define the quantity 
\begin{align}
&\int \dr^2 \bfb e^{i \bfp \cdot \bfb}\Bigg[\exp{\Bigg\{-\frac{\pi}{2}b^2L m_D^3 \ln \frac{2\sqrt{e}}{m_Dbe^{\gamma_E}}\Bigg\}} -1 \Bigg] \nonumber \\
&=  2\pi \int \dr b\, b\, J_0(b|\bfp|) \Bigg[\exp{\Bigg\{-\frac{\pi}{2}b^2L m_D^3 \ln \frac{2\sqrt{e}}{m_Dbe^{\gamma_E}}\Bigg\}}\Bigg] -(2\pi)^2 \delta^2(\bfp) \nonumber \\
& \equiv  (2\pi)^2\left(P(|\bfp|,L) - \delta^2(\bfp)\right).    
\label{eq:ProbB}
\end{align}
For numerical estimates it suffices to set the upper limit for the $b$ integral to 1 GeV$^{-1}$ due to the rapidly decaying exponential factor which we also plotted in Fig.~\ref{fig:Ib}. 
We plot the function $P(|\bfp|) $ as a function of momentum $|\bfp|$ in Fig. \ref{fig:Pp} for medium length $L=4$ fm and two different values of the Debye mass. This function can be interpreted as a probability distribution for the total transverse momentum exchanged with the medium which obeys 
\bea 
\int \dr^2 \bfp\, P(|\bfp|, L) = 1.
\eea
We can now compute the average value of $|\bfp|$ which for our rough estimate of the non-perturbative physics yields a value $\langle |\bfp| \rangle \sim 10$ for L = 4 fm and $m_D = 0.7$ GeV which is a perturbative scale far above the scale $m_D$. Therefore, we conclude that the multiple interactions of the jet partons with the medium ones leads to the emergence of a perturbative scale $Q_{\med}$ of the order of a few GeV. The exact value depends on non-perturbative physics of the medium and its interaction with the cs gluon as well as the medium size. 

With this expression in hand we can rewrite the single subjet function for vacuum broadening term as
\bea
{\cal S}_{1,B}(\epsilon_L, \mu \sim \epsilon_L R)
& =& \frac{\alpha_s(N_c^2-1)}{4\pi^2}\mu^{2\epsilon}  \int\frac{{\rm d}q^-}{q^-}\Big[\delta(\epsilon_L)-\delta(q^--\epsilon_L)\Big]\int {\rm d}^{d}\bfq \Theta\left(|\bfq|- \frac{q^-R}{2}\right)\nonumber \\
&& \Bigg[ \int  \frac{ \dr^2\bfp\,P(|\bfp|, L)}{(\bfq+ \bfp)^2} - \frac{1}{\bfq^2}\Bigg].
\label{eq:vacbroad}
\eea
To further simplify this expression we can now perform the angular integral over $\bfp$ which leads to 
\bea 
I_q = \int \dr^2\bfp \frac{P(|\bfp|, L)}{(\bfq+ \bfp)^2} =  2\pi \int \dr p\,  \frac{p\,P(p, L)}{|p^2-q^2|},
\label{eq:iq1}
\eea
where we have defined $p, q \equiv |\bfp|, |\bfq|$. Note that the above result is divergent as $ p \rightarrow q$. In the presence of the medium this divergence would be regulated with the Debye screening mass $m_D$. Consequently, we rewrite the above expression as 
\bea
I_q =  2\pi \int \dr p \frac{p\,P(p, L)}{\sqrt{(p+q)^2+m_D^2}\sqrt{(p-q)^2+m_D^2}}.
\label{eq:iq}
\eea
Combining Eqs.~\ref{eq:iq} and \ref{eq:vacbroad} we can rewrite the vacuum broadening term for the single subjet function with multiple scatterings as 
\bea
{\cal S}_{1,B}(\epsilon_L, \mu \sim \epsilon_L R)
& =& \frac{\alpha_s(N_c^2-1)}{4\pi^2}\mu^{2\epsilon}  \int\frac{{\rm d}q^-}{q^-}\Big[\delta(\epsilon_L)-\delta(q^--\epsilon_L)\Big]\int {\rm d}^{d}\bfq \Theta\left(|\bfq|- \frac{q^-R}{2}\right)F(q, L)\nonumber \\
\eea
where we have defined 
\bea 
F(q, L) = I_q - \frac{1}{\bfq^2+ m_D^2}.
\label{eq:Fq}
\eea
\begin{figure}
\centering
\includegraphics[width=0.7\linewidth]{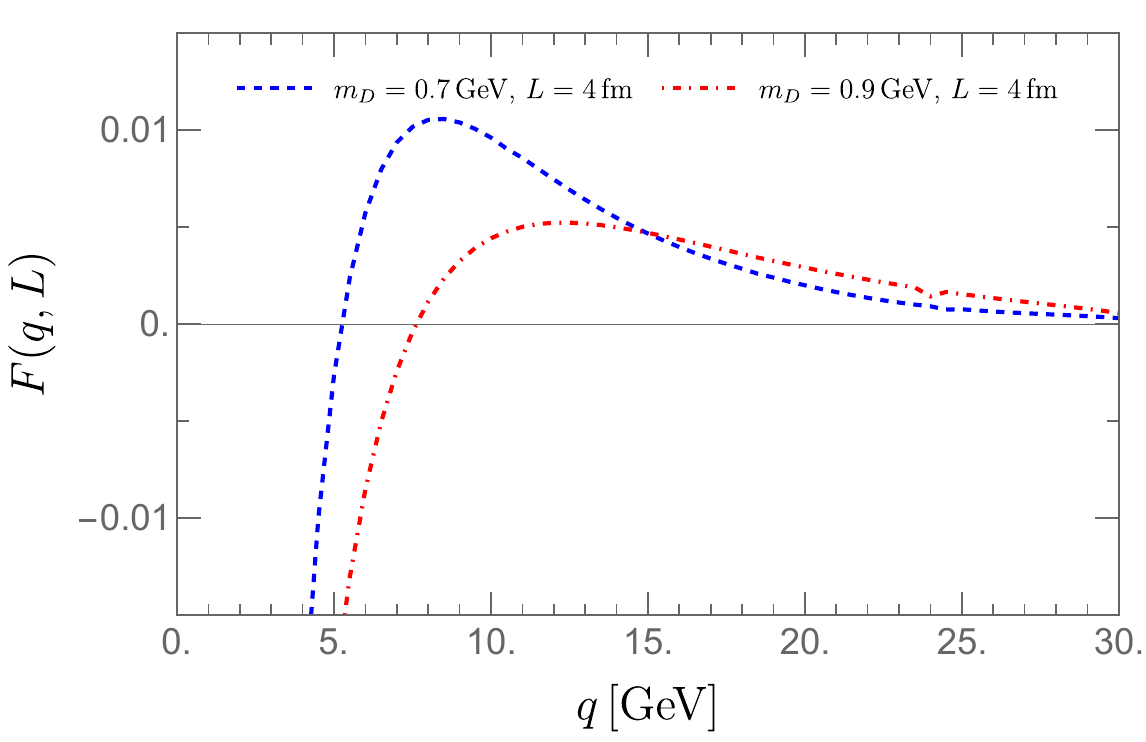}
\caption{The distribution $F(q, L)$ defined in Eq.~\ref{eq:Fq} as a function of transverse momentum $|\bfq|$ for $L=4$ fm and two different values of the screening mass. The peak of the distribution gives a rough estimate for the scale $Q_{\med}$.}
\label{fig:Fq}
\end{figure}
We show the variation of $F(q,L)$ for $L=4$ fm and two different value of screening mass, i.e., $m_D=0.7,0.9$ GeV as a function of transverse momentum $|\bfq|$ in Fig.\ref{fig:Fq}. Moreover, the peak of the distribution provides an estimate for the emergent perturbative scale due to multiple scatterings.  In the absence of the measurement function we numerically verify that 
\bea
\int \dr^2 \bfq F(q, L) = 0, 
\eea
as expected. We can get an estimate of the typical transverse momentum $\bfp_{\text{med}}$ induced by the medium from Fig.~\ref{fig:Fq}. Note that for $\bfq \gg Q_{\text{med}}$, the distribution $F(q,L)$ rapidly decays and eventually approaches to zero.  
 For the case of Fig.~\ref{fig:Fq} this gives us a value $Q_{\text{med}} \sim 10$ GeV which is roughly the peak of the distribution and agrees with our estimate based on the distribution function $P(p, L)$ . 
This then determines the scaling for $|\bfq| \sim Q_{\text{med}} $, the energy $\epsilon_L \sim Q_{\text{med}}/R$ and therefore fixes completely the energy for the collinear soft mode. 
The result after doing the integral over $\bfq$ would therefore depend on two scales $Q_{\text{med}}$ and $m_D$. Given that the two scales appear to be widely separated, we can consider two distinct regimes for $\bfq$. 
To see this, we first shift $\bfq \rightarrow -\bfq - \bfp$ in Eq.~\ref{eq:iq1} and consider the integral over $\bfq$ first to define
\begin{equation}
G(q^-R/2, L) =\int \dr^2\bfp \int {\rm d}^{d}\bfq \Theta\left(|\bfq+\bfp|- \frac{q^-R}{2}\right)  \frac{P(|\bfp|, L)}{\bfq^2+m_D^2} 
- \int {\rm d}^{d}\bfq \Theta\left(|\bfq|- \frac{q^-R}{2}\right)  \frac{1}{\bfq^2+m_D^2}.   
\label{eq:Gq}
\end{equation}
We note that there is no UV divergence since for $\bfq \gg \bfp $, the result goes to zero. 
First for further simplifications, we consider two scalings for $\bfq$. For $\bfq \sim \bfp \sim Q_{\med} \gg m_D$, which we write as the ``hard function" at the scale $Q_{\text{med}}$ and define
 \bea 
 G_H &=&\int d^2\bfp \int {\rm d}^{2-2\epsilon}\bfq \Theta\left(|\bfq+\bfp|- \frac{q^-R}{2}\right)  \frac{P(|\bfp|, L)}{\bfq^2} - \int {\rm d}^{2-2\epsilon}\bfq \Theta\left(|\bfq|- \frac{q^-R}{2}\right)  \frac{1}{\bfq^2},
 \eea
which has an IR divergence in the first term which we regulate using dimensional regularization.  We note that given the scaling for $q^-$ the second term does not have an IR divergence. 
Second we can take the limit $\bfq \rightarrow m_D$ which gives us the  ``IR function" at the scale $m_D$ 
\bea
 G_{IR} &=&\int d^2\bfp \int {\rm d}^{2-2\epsilon}\bfq \Theta\left(|\bfp|- \frac{q^-R}{2}\right)  \frac{P(|\bfp|, L)}{\bfq^2+m_D^2} \nonumber \\
 &= &\Bigg[\int d^2\bfp P(|\bfp|, L) \Theta\left(|\bfp|- \frac{q^-R}{2}\right) \Bigg]\int \frac{{\rm d}^{2-2\epsilon}\bfq} {\bfq^2+m_D^2}. 
\eea
Note that this function now has a UV divergence which will cancel against the IR divergence from the hard function. Intuitively we can understand this IR function as arising from the scenario when the cs gluon from the vacuum Wilson line is emitted with a transverse momentum $m_D \ll Q_{\text{med}}$ which then interacts with the medium and acquires transverse momentum $\bfp \sim Q_{\text{med}}$. 
The full result will therefore produce a logarithmic correction $ \sim \ln Q_{\text{med}}/m_D$. 

In principle, both these regions contribute at leading power to our observable and therefore to completely isolate the non-perturbative physics, we need to do another step of matching from the scale $Q_{\med}$ to $m_D$. However, there is one other aspect we need to account for before doing this factorization, namely quantum interference effects such as the LPM. We see that the lifetime of the gluons with $\bfq \sim m_D$ and $q^- \sim \epsilon_L$ will scale as $t_f \sim \epsilon_L/m_D^2 \sim Q_{\med}/(m_D^2R)$. Depending on the exact values of the scales, this number can be comparable or even much  larger than the medium size. Therefore, one would expect radiation in this phase space to undergo LPM suppression. This suggests that the finite medium length may form a natural IR cutoff $\bfq_{c} \sim \sqrt{Q_{\med}/(LR)}$ which will compete with the medium cut-off scale $\bfq \sim m_D$. Hence a Markovian approximation is not enough and a full calculation with quantum interference effects is needed to fully understand the nature of non-perturbative physics. We will leave this calculation and the matching at the operator level to a future work.

\begin{figure}
\centering
\includegraphics[width=0.7\linewidth]{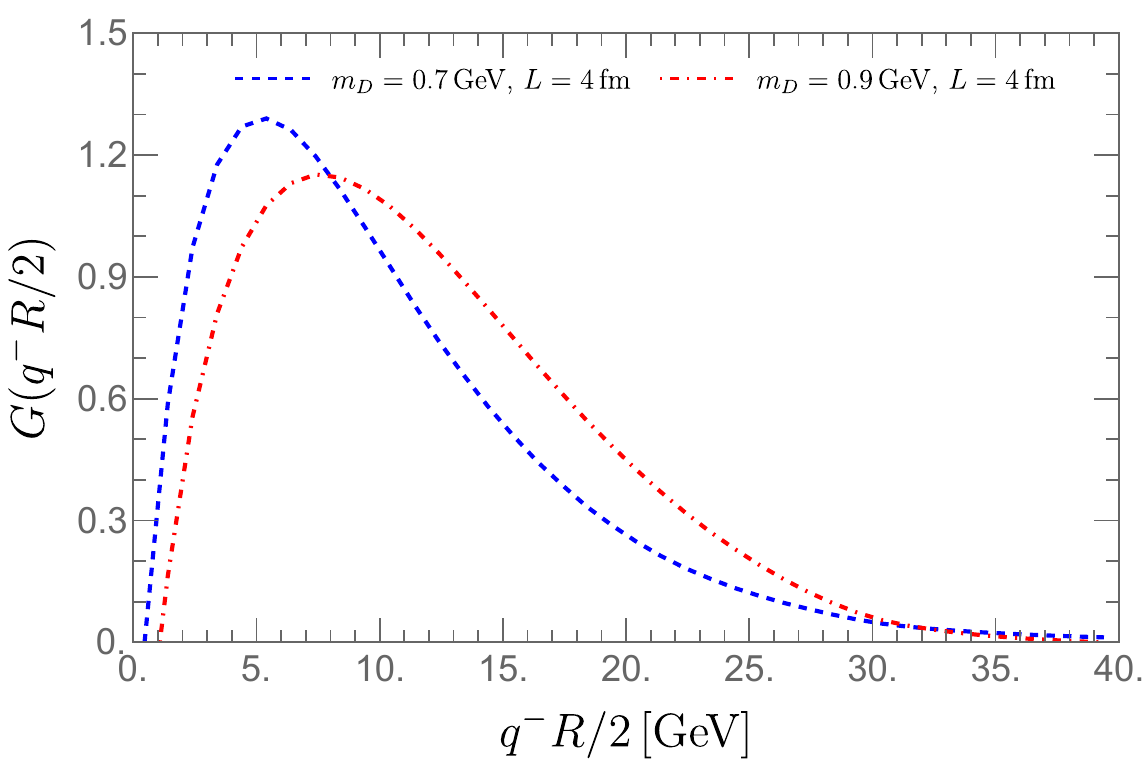}
\caption{The function $G(q^-R/2, L)$ defined in Eq.~\ref{eq:Gq} as a function emitted gluon energy for medium length $L=4$ fm and two different values of the screening mass. The peak of the distribution indicates the existence of the emergent scale $Q_{\rm med}$.}
\label{GD}
\end{figure}

We can now do the integral over $\bfq $ to write our result in terms of a distribution function  $G(q^-R/2, L)$  which we plot in Fig.\ref{GD} for $L=4$ fm and various values of screening mass.
\bea
{\cal S}_{1,B}(\epsilon_L, \mu \sim \epsilon_L R)
& =& \frac{\alpha_s(N_c^2-1)}{4\pi^2}\mu^{2\epsilon}  \int\frac{{\rm d}q^-}{q^-}\Big[\delta(\epsilon_L)-\delta(q^--\epsilon_L)\Big]G(q^-R/2, L),
\eea
We see that the distribution $G(q^-R/2)$ smoothly goes to zero at both small at large values of $q^-$ ensures that the $q^-$ integral has no divergences. We can then write our result in terms of a weighted distribution 
\bea 
{\cal S}_{1,B}(\epsilon_L, \mu \sim \epsilon_L R)
& =& \frac{\alpha_s(N_c^2-1)}{4\pi^2}\Bigg[ \frac{G(\epsilon_LR/2, L)}{\epsilon_L}\Bigg]_+,
\eea
where 
\bea
\int d\epsilon_L f(\epsilon_L)\Bigg[ \frac{G(\epsilon_LR/2, L)}{\epsilon_L}\Bigg]_+ = \int d\epsilon_L \Bigg[ \frac{G(\epsilon_LR/2, L)}{\epsilon_L}\Bigg]\left( f(0) -f(\epsilon_L) \right).
\eea
\subsection{Analysis for medium induced radiation}
We can perform a similar analysis for the case of medium induced radiation. The result for multiple scatterings in the Markovian approximation reads (From Eq.~\ref{eq:S1M})
\begin{align}
&{\cal S}_{1,M}(\epsilon_L, \mu) = \frac{2\alpha_s}{\pi^2}\mu^{2\epsilon}  \int\frac{{\rm d}q^-}{q^-}\Big[\delta(\epsilon_L)-\delta(q^--\epsilon_L)\Big]\int \frac{\dr^2\bfp}{(2\pi)^2}\int \dr^2 \bfb\, e^{i \bfp \cdot \bfb}\int \frac{{\rm d}^{d}\bfq  }{\bfq^2 }\Theta\left(|\bfq +\bfp|- \frac{q^-R}{2}\right)\nonumber \\
&\qquad\,\quad\Bigg[\frac{\int \frac{\dr^2 \bfu}{(\bfu+\bfq)^2} \varphi_R(\bfu, \nu_{\cs} )}{\int\frac{\dr^2\bfl}{\bfl^2}\left(1-e^{-i \bfl \cdot \bfb}\right)\varphi_R(\bfl, \nu_{\cs})}\Bigg] \Bigg[\exp{\Big\{-L\int  \frac{d^2\bfk}{\bfk^2  }\left(1-e^{-i \bfk \cdot \bfb} \right)\varphi_R(\bfk, \nu_{\cs})\Big\}}-1 \Bigg]. 
\end{align}
The exponential factor that leads to the emergent scale $Q_{\med}$ as shown in the previous section is identical and hence we again have two scales to deal with $Q_{\med}, m_D$. Consequently, we can again consider two possibilities for the scaling of $\bfq$. We note that the natural scaling for $\bfu$ is just $m_D$.  Therefore, as before we can consider two regions 
\begin{itemize}
    \item{$\bfq \sim m_D$}
\begin{align}
&{\cal S}^{IR}_{1,M}(\epsilon_L, \mu) = \frac{2\alpha_s}{\pi^2}\mu^{2\epsilon}  \int\frac{{\rm d}q^-}{q^-}\Big[\delta(\epsilon_L)-\delta(q^--\epsilon_L)\Big]\nonumber\\
& \times  \int \frac{\dr^2\bfp}{(2\pi)^2}\int \dr^2 \bfb\, e^{i \bfp \cdot \bfb}\frac{\Bigg[\exp{\Big\{-L\int  \frac{d^2\bfk}{\bfk^2  }\left(1-e^{-i \bfk \cdot \bfb} \right)\varphi_R(\bfk, \nu_{\cs})\Big\}}-1 \Bigg] }{\int\frac{\dr^2\bfl}{\bfl^2}\left(1-e^{-i \bfl \cdot \bfb}\right)\varphi_R(\bfl, \nu_{\cs})}\Theta\left(|\bfp|- \frac{q^-R}{2}\right)\nonumber \\
& \times \int \frac{{\rm d}^{d}\bfq  }{\bfq^2 }\int \frac{\dr^2 \bfu}{(\bfu+\bfq)^2} \varphi_R(\bfu, \nu_{\cs} ).
\end{align}
We note that the result can be written in terms of the same universal broadening factor $P(|\bfp|, x^-)$ defined in the previous section as 
\begin{align}
&{\cal S}^{IR}_{1,M}(\epsilon_L, \mu) =  \int\frac{{\rm d}q^-}{q^-}\Big[\delta(\epsilon_L)-\delta(q^--\epsilon_L)\Big] \Bigg[ \int_0^L \dr x^- \int \frac{\dr^2\bfp}{(2\pi)^2}P(|\bfp|,x^-)\Theta\left(|\bfp|- \frac{q^-R}{2}\right)\Bigg] \nonumber\\ 
& \times \frac{2\alpha_s}{\pi^2}\mu^{2\epsilon} \int \frac{{\rm d}^{d}\bfq  }{\bfq^2 +m_D^2}\int \frac{d^2 \bfu}{(\bfu+\bfq)^2 +m_D^2} \varphi_R(\bfu, \nu_{\cs} ).
\end{align}
Here we see that the integral over $\bfq$ is finite, i.e. the IR physics is regulated by $m_D$ and there is no UV divergence. 
\item{$\bfq \sim \bfp $}
\begin{align}
&{\cal S}_{1,M}^{UV}(\epsilon_L, \mu) = \frac{2\alpha_s}{\pi^2}\mu^{2\epsilon}  \int\frac{{\rm d}q^-}{q^-}\Big[\delta(\epsilon_L)-\delta(q^--\epsilon_L)\Big]\int \frac{\dr^2\bfp}{(2\pi)^2}\int \frac{{\rm d}^{d}\bfq}{(\bfq^2)^2 }\Theta\left(|\bfq +\bfp|- \frac{q^-R}{2}\right)\nonumber \\
&\qquad\,\quad\int \dr^2 \bfu\varphi_R(\bfu, \nu_{\cs} ) \int \dr x^- P(|\bfp|,x^-),
\end{align}
we note that compared to the IR contribution this scales as $m_D^2/\bfp^2 \sim m_D^2/Q_{\med}^2$ and hence is a power suppressed contribution which can be ignored. 
\end{itemize}
Therefore we conclude that only the IR regime, i.e. when the medium induced gluon is emitted with  an initial transverse momentum $\bfq \sim m_D$ contributes at leading power to the energy loss observable. 
\section{Matching from $Q_{\text{med}}$ to $m_D$}
\label{sec:Match2}
In the previous section, with some assumptions about the scaling of non-perturbative physics, we have shown the existence of an emergent transverse scale $Q_{\text{med}}$ which is well separated from the jet scale $p_TR$ as well as from the non-perturbative medium scale $m_{\d}$ (or $T$) and leads to the scale hierarchy $p_TR \gg Q_{\text{med}} \gg m_D$.  We note here that $Q_{\med}$ is not a single fixed scale but is distributed over a range of transverse scales with some probability, however, for for a consistent factorized framework the important conclusion is that it obeys this hierarchy.
Note that this appears when we account for multiple scatterings between jet partons and the medium.
The factorization formula derived so far in Eq. \ref{eq:J-S-fact}  separates the jet dynamics at scale $p_TR$ from the  IR physics at scale $Q_{\text{med}}$ which includes the physics of all the scale down to $m_D$. In light of the new hierarchy arising from the multiple scatterings, it is clear that to fully separate out the perturbative physics from the non-perturbative one, we need to further separate the physics at $Q_{\text{med}}$ from $m_D$ which we will discuss below. Although we will not finish the task completely in this paper, we will take a significant step in this direction. 

Based on the analysis of the leading order results obtained in the previous section, the energy of the collinear soft mode is fixed to be $Q_{\text{med}}/R$ where $Q_{\text{med}}$ is net transverse momentum gained by the cs gluon through multiple interactions with the medium. This means that although the final transverse momentum of the gluon after it moves out of the medium will scale as $Q_{\med}$, at the time of its emission it can have a much smaller transverse momentum $\sim  m_D$. We have already seen how this appears through an explicit calculation for both vacuum and medium induced radiation in the previous section.
 Therefore, \textit{at the source} the transverse momentum and hence virtuality of the radiation can have two distinct scalings leading us to define two modes, a collinear soft(cs) mode that is already part of our EFT framework 
\bea
p_{\cs} \sim \frac{Q_{\text{med}}}{R}\left( 1,R^2 , R \right),
\eea
and an ultra-collinear soft (ucs) mode, involving non-perturbative scale of the medium, with the momentum scaling
\bea
p_{\ucs} \sim \frac{Q_{\text{med}}}{R}\left( 1,  \frac{m_{\d}^2R^4}{Q_{\med}^2}, \frac{m_{\d}R}{Q_{\text{med}}}\right). 
\eea
Note that the ucs mode has the same energy as that of the cs mode but has a much lower virtuality $p_{\rm ucs}^2 \sim m_D^2$. After emission at the source, either through vacuum or medium-induced emission, both these types of gluons will undergo multiple scattering off the medium partons. During their evolution and interactions in the medium, these modes will acquire the total transverse momentum  $|\bfp| \sim Q_{\med}$ with a distribution that has the form
\bea
P(\bfp, L) = \int \dr^2\bfb\, e^{i \bfp \cdot \bfb}\Big[\exp{\Big\{-L \int \frac{\dr^2\bfk}{\bfk^2}(1- e^{-i \bfk \cdot b})\varphi_R(\bfk, \nu_{cs}\Big\}} \Big].
\label{eq:distribution}
\eea
Note that this distribution function depends on both $Q_{\med} \sim 1/b$  as well as medium scale $m_D$. Hence, owing to the hierarchy (with multiple scattering) a separation of scale is also required within this distribution function.
In this paper, we will consider the problem of factorizing this distribution function and leave the issue of separating the ucs mode from the cs in a future work.
We can therefore consider two regimes for the transverse momentum transfer to the jet in a single interaction, 
\begin{itemize}
\item{$\bfk \sim m_D \ll 1/b $}. In this case, the transverse momentum distribution simplifies to  
\bea
P_{\text{soft}}(\bfp, L; \mu)=  P(\bfp, L) \xrightarrow[\bfk \sim m_D]{} \int \dr^2\bfb\, e^{i \bfp \cdot \bfb}\Big[\exp{\Big\{- 2\pi L\, \bfb^2\int \dr^2\bfk  \varphi_R(\bfk, \nu_{cs}\Big\}} \Big].
\eea
Note that this case is similar to single scattering limit where momentum transfer in each interaction is approximately same as $m_D$. Therefore, the medium function $\varphi_R$ obeys full BFKL evolution equation. The jet in this regime therefore probes the small $x$ dynamics of the medium with $ x \sim m_D/\epsilon_L$. We also note that this object has a UV divergence which appears in $\bfk$ integral as $\bfk \rightarrow \infty$. Hence $P_{\text{soft}}$ explicitly depends on a renormalization scale $\mu$.

While the definition of a local correlator is standard for a single color correlated state such as a hadron, it is less clear how to define a such an object for an extended medium such as a large nucleus and even more obscure for quark gluon plasma. A useful observation which has been used earlier in Ref.~\cite{Kumar:2019uvu} is that any correlator such as $\varphi$, which is a measure of a local parton density  may be meaningfully defined on a region in which the partons are color correlated.  This means that we need to carve up the medium into local regions with a typical size decided by the color screening length in the medium. For the rest of this paper we will refer to these coherent regions as nucleons of the medium.
We have seen in the earlier section that the assumption of independent scatterings, i.e., the successive interactions of the jet with the medium happen with color uncorrelated partons, is an excellent approximation. Therefore, the successive interactions of the jet with the medium happen with distinct Nucleons. Therefore we introduce a Nucleon density function $\rho$, which is a constant for a simple model of homogeneous medium. This allows to write
\bea
P_{\text{soft}}(\bfp, L; \mu)=  \int \dr^2\bfb e^{i \bfp \cdot \bfb}\Big[\exp{\Big\{- 2\pi L \rho \, \bfb^2\int \dr^2\bfk  \frac{1}{P^+}\varphi_R(\bfk, \nu_{cs}\Big\}} \Big].
\eea
where $\varphi$ is now defined on a single Nucleon state $|N\rangle$. The $+$ component of momentum of this plane wave state is given by $P^+$.
\begin{align}
&\varphi(\bfk,\mu,\nu)= \frac{1}{\bfk^2}\frac{1}{N_c^2-1}\int \frac{{\rm d}k^-}{2\pi}\int {\rm d}^4r\, e^{-i \bfk \cdot \bfr+ik^-r^+}\text{Tr}\Big[e^{-i\int {\rm d}t H_s(t) }\mathcal{O}^A_{\s}(r)|N \rangle \langle N| \mathcal{O}^A_{\s}(0)e^{i\int {\rm d}t H_s(t) }\Big].\,\qquad
\label{eq:varphiN}
\end{align}
We can combine $\rho L$ as $\rho_A$ which is the longitudinal boost invariant Nucleon density per unit transverse area of the medium.  
We give a detailed derivation of this equation in Appendix \ref{app:Nucleon}. It is easy to generalize this for medium inhomogeneity along the direction of propagation of the jet and write 
\bea
P_{\text{soft}}(\bfp, L; \mu)= \int \dr^2\bfb e^{i \bfp \cdot \bfb}\Big[\exp{\Big\{- 2\pi \int_0^L \dr x^- \rho(x^-) \, \bfb^2\int \dr^2\bfk  \frac{1}{P^+}\varphi_R(\bfk, \nu_{cs}\Big\}} \Big].
\eea
  \item{$\bfk \sim Q_{\med} \gg m_D $}.
This limit corresponds to a large momentum transfer in a single interaction and will therefore be the tail of the transverse momentum distribution. 
We note that in this regime, 
\bea
x = \frac{ \text{probe virtuality}}{\text{c.o.m. energy}} \sim \frac{Q_{\med}^2}{m_D Q_{\med}/R} = \frac{Q_{\med} R}{m_D} \sim 1.
\eea
Hence, this region no longer falls in the forward scattering scenario but in fact requires a twist expansion with a full Parton Distribution Function structure. 
The other implication is that the transfer of energy between the medium partons and the cs partons, i.e., collisional energy loss of the jet cannot be ignored. This hard scattering is a (relatively) large angle scattering referred to as Moliere scattering which was explored for single interaction in \cite{DEramo:2018eoy}. Due to the large momentum transfer the value of $\alpha_s$ is smaller compared to the regime $\bfk \sim m_D$ but we see that strictly by power counting, it contributes at the same order as the forward scattering regime. Hence it now becomes necessary to revisit our framework and include this regime if we are to be consistent with our EFT expansion. In this paper, we will not attempt a rigorous derivation of this regime but  will simply put in a possible structure consistent with our intuition, while still ignoring energy loss. We will leave a detailed analysis including incorporating collisional energy loss in the EFT for the future. 

We therefore propose to match the function $\varphi_R(\bfk, \nu_{\cs})$ to a new operator integrating out physics at the scale $\bfk$. Given that this regime corresponds to hard scattering, we can expect the operator to be some type of Parton Distribution Function(PDF) in the nuclear medium. This allows us to write\footnote{We stress that this does not capture the full hard scattering cross section since we are ignoring collisional energy loss}
\bea
P_{\text{hard}}(\bfp, L; \mu) &=& P(\bfp, L) \xrightarrow[\bfk \sim 1/b]{}  \int \dr^2 \bfb\, e^{i \bfp \cdot \bfb}\Bigg[\text{exp}\{\Big\{- \int \frac{\dr^2\bfk}{\bfk^2}(1- e^{-i \bfk \cdot b})\nonumber\\
&& \int_0^L \dr x^-\rho(x^-) \int_0^1 \frac{\dr \xi}{\xi}C(\bfk, x/\xi;\mu) Y(\xi)\Big\}\}\Bigg],  
\eea
 where 
$x = \bfk^2/(P^+\epsilon_L)$ and
\bea
Y= \int_{\xi}^{1}\frac{\dr z}{z} P_{gq}(z)f_q(\xi/z, \mu)
\eea
where for simplicity we have assumed the medium to be populated by quarks. The case for gluons in the medium can be handled by adding an analogous term .
At tree level 
\bea
C(\bfk; x/\xi;\mu) = \frac{x}{C_F}(8 \pi \alpha_s(\mu))^2\frac{1}{\bfk^2}\delta\left(1- \frac{x}{\xi}\right),
\eea
and $P_{gq}$ are the Altarelli-Parelli splitting function. 
The quark PDF in the medium is defined on a single Nucleon plane wave state $|N\rangle$ introduced earlier and has the usual definition.
\bea
 f_q(x) =\int \frac{\dr y}{2\pi} e^{-i y x P^+}\text{Tr}\Big[\chi_n(y) \frac{\slashed{\bar n}}{2}|N\rangle \langle N| \bar \chi_n(0) \Big],
\eea
 where $y$ is a separation along the direction of jet propagation $x^-$.
The PDF obeys the DGLAP evolution equation and we can do the resummation from some perturbative scale $Q_0$ to $\mu \sim k_{\perp}$. 
The non-perturbative physics is then encoded in the boundary condition, namely $Y(\xi, \mu= Q_0^2)$. 
We see that the function $P_{\text{hard}}(\bfp)$ has an IR divergence as $\bfk \rightarrow 0$ which we also regulate using dim. reg. This IR divergence exactly compensates the UV divergence for the $P_{\text{soft}}(\bfp)$ so that the full result is finite. 
\end{itemize}
Putting everything together, we can then write 
\bea
P(\bfp, L) &=& \int \dr^2\bfb\, e^{i \bfp \cdot \bfb}\Big[\exp{\Big\{-\int \dr x^- \frac{\rho(x^-)}{P^+} 2\pi |\bfb|^2 \Phi (R; \mu)\Big\}}\nonumber \\
& \times & \exp{\Big\{-\int \dr x^- \rho(x^-)  \int \frac{\dr^2\bfk}{\bfk^2}(1- e^{-i \bfk \cdot b})\int_0^1 \frac{\dr\xi}{\xi}C(\bfk, x/\xi;\mu)  Y(\xi) \Big\}}\Big]+ O\left(\frac{m_D^2}{Q_{\med}^2}\right) \nonumber \\
\label{eq:NP}
\eea

The non-perturbative physics is therefore encoded in two objects $\Phi (R; \mu)$ and $Y$ where 
\bea 
\Phi (R; \mu)= \int \dr^2\bfk\,  \varphi_R(\bfk, \nu_{cs}).
\eea
These are two universal objects that describe the strongly coupled physics of the medium. We note that through the dependence in $\nu_{cs}$, $x$ they depend on the angular measurement made on the jet, i.e., the radius R.

We see that the evolution of a high energy parton in the medium encompasses both the large $x$ and small $x$ limit of QCD. 
This is interesting since even though we started off with the assumption of only forward (and hence small $x$) as the dominant interaction mechanism, the emergence of the scale $Q_{\med}$ also enforces large $x$ physics.
We expect that in the region of overlap $ 1/b \gg \bfk \gg m_D$, the two pieces should be identical to each other.  This is the Double Logarithmic Approximation(DLA) where the BFKL and DGLAP evolution are identical. So we expect 
\bea
\lim_{\bfk \rightarrow \infty} P_{\text{soft}}(\bfp, L; \mu)= \lim_{\bfk \rightarrow 0} P_{\text{hard}}(\bfp, L; \mu)
\eea
Here we test this explicitly which is also a check on our factorization. 

\section{ The overlap regime and $\hat q$}

To check the overlap we do an explicit perturbative calculation assuming the medium Nucleon to be a single quark with momentum p. It suffices to look at the arguments of the exponents. 
In this case  the soft or $\bfk \rightarrow 0$ limit of $P_{\text{hard}}$ corresponds to the $x \rightarrow 0$ limit. 
\begin{equation}
\lim_{\bfk \rightarrow 0}P_{\text{hard}}(\bfp, L) = \int \dr^2\bfb e^{i \bfp \cdot \bfb}\Big[\exp{\Big\{-|\bfb|^2\int \dr x^- \rho(x^-) \int \dr^2\bfk \lim_{x\rightarrow 0}\int_0^1 \frac{\dr \xi}{\xi} C(k_{\perp}, x /\xi;\mu)Y(\xi)\Big\}}\Big]    
\end{equation}
Plugging in the tree level value for the coefficient $C$, we have 
\begin{equation}
\lim_{\bfk \rightarrow 0}P_{\text{hard}}(\bfp, L)=\int \dr^2\bfb e^{i \bfp \cdot \bfb}\Big[\exp{\Big\{-|\bfb|^2\int \dr x^- \rho(x^-)\frac{(8\pi \alpha_s)^2}{C_F}\int \frac{\dr^2\bfk }{\bfk^2}\lim_{x\rightarrow 0}xY(x, \mu \sim \bfk)\Big\}}\Big].    
\end{equation}
Since 
\bea
Y^{(0)}(x, \mu ) = \int_x^1 \frac{\dr z}{z}P_{gq}(z) f_q\left(\frac{x}{z}, \mu \right),
\eea
where for simplicity we only consider a medium composed of quarks.
Consequently 
\bea
\lim_{\bfk \rightarrow 0}P_{\text{hard}}(\bfp, L) &=& \int d^2\bfb e^{i \bfp \cdot \bfb}\Big[\exp{\Big\{-|\bfb|^2\int dx^- \rho(x^-)\frac{(8\pi \alpha_s)^2}{C_F}\int \frac{d^2\bfk}{\bfk^2} \lim_{x\rightarrow 0} x\int_x^1 \frac{dz}{z}P_{gq}(z)\tilde f_q\left(\frac{x}{z}, \mu \sim \bfk \right)\Big\}}\Big] \nonumber \\
\eea
For comparison we explicitly evaluate the PDF at tree level in a medium Nucleon composed a single quark with momentum p 
\bea
\lim_{\bfk \rightarrow 0}P_{\text{hard}}(\bfp, L) &=&  \int \dr^2\bfb\, e^{i \bfp \cdot \bfb}\Big[\exp{\Big\{-|\bfb|^2\int \dr x^- \rho(x^-)(8\pi \alpha_s)^2\int \frac{\dr^2 \bfk}{\bfk^2} \Big\}}\Big] 
\label{eq:Psh}
\eea
We can equivalently write this in terms of the small x limit of the gluon PDF
\bea
\lim_{\bfk \rightarrow 0}P_{\text{hard}}(\bfp, L) &=& \int \dr^2\bfb e^{i \bfp \cdot \bfb}\Big[\exp{\Big\{-|\bfb|^2\int \dr x^- \rho(x^-) (8\pi \alpha_s)^2 \left(\frac{\pi}{\alpha_s C_F}\right) \lim_{x \rightarrow 0}x f_g(x, \mu)\Big\}}\Big] 
\eea
At higher orders we know that in the small x limit of the gluon PDF obeys the Double Logarithmic Approximation  of the DGLAP equation. 
Parametrically the integral over $\bfk$ would range from $\bfk \in \{\mu, 1/b \}$ which would cut-off the UV and IR singularities in this overlap region. 
Likewise we can consider the $\bfk \rightarrow \infty$ limit of $P_{\text{soft}}$
\bea
\lim_{\bfk \rightarrow \infty}P_{\text{soft}}(\bfk, L) = \int \dr^2\bfb\, e^{i \bfp \cdot \bfb}\Big[\exp{\Big\{-2\pi |\bfb|^2 \int \dr x^- \rho(x^-)\int \dr^2\bfk  \lim_{\bfk \rightarrow \infty} \frac{1}{P^+}\varphi_R(\bfk, \nu_{cs}\Big\}} \Big]
\eea
 Using the definition for $\varphi$ at tree level 
\bea
\varphi (\bfk) &=& \frac{(8\pi \alpha_s)^2}{\bfk^2}\frac{1}{N_c^2-1}\text{Tr} \Big[ \delta^2(\bfk -\mathcal{P}_{\perp})\delta(\mathcal{P}^+)\Big[\bar \chi_s T^A \chi\Big]  |p \rangle \langle p | \bar \chi T^A \chi_s \Big] \nonumber \\
&=& \frac{P^+}{\bfk^2}
\eea
so that the tree level result becomes 
\bea
\lim_{\bfk \rightarrow \infty}P_{\text{soft}} = \int \dr^2\bfb\, e^{i \bfp \cdot \bfb}\Big[\exp{\Big\{- 2\pi |\bfb|^2(8\pi \alpha_s)^2 \int \dr x^- \rho(x^-)\int \frac{\dr^2\bfk}{\bfk^2} \Big\}} \Big]
\eea
which then agrees in form  with the tree level result in Eq.\ref{eq:Psh}. In this case $\bfk \in \{ m_D, \mu \}$ so that the UV divergence cancels exactly with the IR divergence in Eq.\ref{eq:Psh}.
 We already know that in the limit $\bfk \rightarrow \infty$, the BFKL resummation reduces to the DLA result so that the two results also agree in form at all orders in $\alpha_s$ in this limit which is a non-trivial check on our factorization. 

If we only keep the overlap region then we can roughly set the limits for $\bfk$ integral to be $\bfk \in \{ m_D, 1/b  \sim 1/Q_{\med}\}$ which would then give us from Eq.\ref{eq:Psh} the conventional definition of the jet transport parameter $\hat q$ if we assume a homogeneous medium
\bea
P_{\text{overlap}}&=& \int \dr^2\bfb e^{i \bfp \cdot \bfb}\Big[\exp{\Big\{-|\bfb|^2\int_0^L \dr x^- \rho \frac{\pi (8\pi \alpha_s)^2}{\alpha_sC_F}\lim_{x \rightarrow 0}x f_g(x, \mu \sim Q_{\med})\Big\}}\Big] \nonumber \\
&\equiv &\int \dr^2\bfb e^{i \bfp \cdot \bfb}\Big[\exp{\Big\{-|\bfb|^2 L \hat q\Big\}}\Big]
\eea 
where $\hat q =\frac{\pi (8\pi \alpha_s)^2}{\alpha_sC_F} \rho \lim_{x \rightarrow 0}x f_g(x = m_D/(\epsilon_L P^+), \mu \sim Q_{\med})$. 
However, we see that the full non-perturbative physics consistent with our power counting is given by Eq.~\ref{eq:NP} and therefore contains both hard and forward scattering probes of the medium. 

\section{Conclusion and Outlook}
\label{sec:Con}
Understanding the nature of non-perturbative physics for jet propagation is one of the most important questions that needs to be answered if we are to make any progress in quantitatively describing jet observables in heavy ion collisions.
This requires a precise parameterization of the non-perturbative physics in terms of well defined operators which will enable us to test the universality of the non-perturbative physics. A crucial step towards this goal is a factorization formula that cleanly separates physics at distinct scales to all orders in $\alpha_s$. A significant step towards this goal was taken through a factorization formula presented in \cite{Mehtar-Tani:2024smp}. However, the presence of emergent scales which can only be see through an explicit computation requires further steps in order to achieve complete factorization. 

In this paper, we explore the consequences of the factorization formula for jet propagation in a large but dense medium to understand whether further factorization is necessary. Starting from a factorization formula for inclusive jet production written in terms of subjet functions, we looked at the computation of the single subjet function which is the only relevant term for an unresolved jet. We showed that the Markovian approximation of multiple scatterings of a gluon sourced by either vacuum evolution or medium induced radiation in this subjet function leads to an emergent transverse momentum scale $Q_{\med}$ that depends on the density and spatial extent of the medium. For the phenomenological properties of the medium encountered in current experiments, this a perturbative scale well separated from the non-perturbative Debye mass scale of the strongly coupled medium. While this scale has been noted in earlier works, here we examine the consequences of this scale on the structure of factorization.
We first show that the broadening of a gluon in the medium can be described by a (semi)universal probability distribution that depends on the angular measurement made on the radiation, in this case the jet radius R. Given the presence of two well separated scales $Q_{\med}$ and $m_D$, this probability distribution can be factored into two pieces at leading order in $m_D/Q_{\med}$. One piece describes forward scattering with the medium and obeys the BFKL evolution, while the other describes hard scattering with the medium and obeys the DGLAP evolution. Thus remarkably the dense medium enforces a hard scattering regime to contribute at the same order in power counting as forward scattering. We show that the conventional definition of the jet transport parameter $\hat q$ is the limiting case of the two pieces in their  region of overlap which is proportional to the small $x$ limit of the gluon PDF and obeys the Double Logarithmic Approximation to the BFKL, DGLAP evolution equations. This leads to the important conclusion that parameterization  of the non-perturbative physics for jet propagation is more complex than is currently used for phenomenological studies.

We also show that there is another source of non-perturbative physics which also needs to be accounted for to completely isolate the physics at $m_D$, which is crucial to understand the universality (or lack thereof ) of the non-perturbative physics. 
This requires us to factorize the physics of gluons \textit{sourced} at two transverse momenta(or virtualities), namely $Q_{\med}$ and $m_D$, which subsequently acquire a transverse momentum $\sim Q_{\med}$ over multiple interactions in the medium.
However for gluons sourced with a transverse momentum $m_D$, the coherence or formation time can be large and hence the Markovian approximation used in this paper breaks down. Therefore to complete this factorization requires us to account for quantum interference effects between successive interactions which will introduce another IR cut-off scale $\sqrt{\text{Gluon energy}/\text{medium length}}$, which suppresses all radiation below this virtuality. This cut-off which may be perturbative would then supersede the non-perturbative cut-off imposed by the Debye mass $m_D$. We leave this calculation for a future work. We also leave open the case of multiple subjets which will be releva0nt whenever the color decoherence angle($\theta_c$) is smaller than the jet radius. A similar non--perturbative analysis and refactorization will also be needed in that case, now with the presence of an additional emergent angular scale $\theta_c$.

\section*{Acknowledgments}
V.V. and B.S.  are supported by startup funds from the University of South Dakota and by the U.S. Department of Energy, EPSCoR program under contract No. DE-SC0025545. 

\appendix
\section*{Appendix}


\section{Correlators in an extended medium} 
\label{app:Nucleon}
We consider a density matrix of the medium $\rho_M$ defined by the normalized state $|\psi \rangle$ as 
\bea
\rho =  |\psi \rangle \langle \psi |, \ \ \ \text{with} \ \ \langle \psi| \psi \rangle = 1
\eea
We now divide the extended medium into its constituent Nucleons which are local regions over which color coherence is maintained. In a QGP medium in its rest frame, this would roughly be a sphere of size $1/m_D$, i.e. the color screening length in the medium. We consider a medium composed of n such identical Nucleons. We can then write our state as 
\bea
|\psi \rangle &= & \frac{1}{\sqrt{V}}\Bigg[\prod_{i=1}^n \int \frac{dP_i^+}{\sqrt{P_i^+}}\int d^2 \bfP_{i}\Bigg]\delta^2(\sum_i\bfP_i) \delta(P_M^+- \sum_i P_i^+)\nonumber \\
&& \psi(P_1^+, .. P_n^+; \bfP_1^{\perp}, ...,\bfP_n^{\perp}) | P_1, P_2, .... P_n\rangle 
\eea
The state has a total longitudinal momentum $P_M^+$ and net zero transverse momentum. The function $\psi$ is the Nuclear wavefunction. The wavefunction ensures that the Nucleons are localized over regions of size $1/m_D$. The volume factor V ensures the normalization of the state to 1. 
We want to compute the correlator $\varphi$ in the density matrix defined by this state. 
\begin{align}
&\varphi(\bfk)= \frac{1}{\bfk^2}\frac{1}{N_c^2-1}\int {\rm d}^2 \bfr\, e^{-i \bfk \cdot \bfr}\text{Tr}\Big[\delta(\mathcal{P}^+)\Big[\mathcal{O}^A_{\s}(\bfr)\Big]|\psi \rangle \langle \psi| \mathcal{O}^A_{\s}(0)\Big].\,\qquad
\end{align}
Plugging in the definition of the state, we have 
\begin{align}
&\varphi(\bfk)= \frac{1}{V}\Bigg[\prod_{i=1}^n \int \frac{dP_i^+}{\sqrt{P_i^+}}\int d^2 \bfP_{i}\Bigg]\delta^2(\sum_i\bfP_i) \delta(P_M^+- \sum_i P_i^+)\psi(P_1^+, .. P_n^+; \bfP_1^{\perp}, ...,\bfP_n^{\perp})\nonumber \\
&\Bigg[\prod_{j=1}^n \int \frac{dQ_j^+}{\sqrt{Q_j^+}}\int d^2 \bfQ_{j}\Bigg]\delta^2(\sum_j\bfQ_i) \delta(P_M^+- \sum_j Q_j^+)\psi^*(Q_1^+, .. Q_n^+; \bfQ_1^{\perp}, ...,\bfQ_n^{\perp})\nonumber \\
&\frac{1}{\bfk^2}\frac{1}{N_c^2-1}\int {\rm d}^2 \bfr\, e^{-i \bfk \cdot \bfr}\text{Tr}\Big[\delta(\mathcal{P}^+)\Big[\mathcal{O}^A_{\s}(\bfr)\Big]| P_1, P_2, .... P_n\rangle  \langle  Q_1, Q_2, .... Q_n| \mathcal{O}^A_{\s}(0)\Big].\,\qquad
\end{align}
The operator acts on the Nucleon localized at location 0, and probes it over a distance $\bfr \sim 1/\bfk \sim 1/m_D$. Therefore it picks out a single Nucleon, say $|P_J\rangle$. The rest of the localized Nucleon states annihilate each other and we are left with 
\begin{align}
&\varphi(\bfk)=  \frac{1}{V}\int \frac{dP_J^+}{P_J^+}\int d^2 \bfP_{J} |\tilde \psi( P_J^+, \bfP_J)|^2\nonumber \\
&\frac{1}{\bfk^2}\frac{1}{N_c^2-1}\int {\rm d}^2 \bfr\, e^{-i \bfk \cdot \bfr}\text{Tr}\Big[\delta(\mathcal{P}^+)\Big[\mathcal{O}^A_{\s}(\bfr)\Big]| P_J\rangle  \langle  P_J| \mathcal{O}^A_{\s}(0)\Big].\,\qquad
\end{align}
where 
\bea
|\tilde \psi( P_J^+, \bfP_J)|^2 =  \Bigg[\prod_{i=1, i\neq J}^n \int \frac{dP_i^+}{\sqrt{P_i^+}}\int d^2 \bfP_{i}\Bigg]|\psi(P_1^+, .. P_n^+; \bfP_1^{\perp}, ...,\bfP_n^{\perp})|^2
\eea
We know that the state $|\psi \rangle $ normalizes to 1  which requires
\bea
\int dP_J^+\int d^2 \bfP_{J} |\tilde \psi( P_J^+, \bfP_J)|^2 = 1
\eea
As a toy illustration, we can consider the Nucleon J to have a wavefunction 
\bea 
\tilde \psi( P_J^+, \bfP_J) = \delta^2(\bfP_{J}) \delta(P_J^+-P^+) 
\eea
which would satisfy this condition yielding 
\begin{align}
&\varphi(\bfk)=  \frac{1}{V} \frac{1}{P+}\int \frac{1}{\bfk^2}\frac{1}{N_c^2-1}\int {\rm d}^2 \bfr\, e^{-i \bfk \cdot \bfr}\text{Tr}\Big[\delta(\mathcal{P}^+)\Big[\mathcal{O}^A_{\s}(\bfr)\Big]| P_J\rangle  \langle  P_J| \mathcal{O}^A_{\s}(0)\Big].\,\qquad
\end{align}
In this scenario we can then interpret the factor $1/V$ as the Nucleon density of the medium $\rho$ which prompts us to write 
\begin{align}
&\varphi(\bfk)= \rho\frac{1}{P+}\int \frac{1}{\bfk^2}\frac{1}{N_c^2-1}\int {\rm d}^2 \bfr\, e^{-i \bfk \cdot \bfr}\text{Tr}\Big[\delta(\mathcal{P}^+)\Big[\mathcal{O}^A_{\s}(\bfr)\Big]|N\rangle  \langle N | \mathcal{O}^A_{\s}(0)\Big].\,\qquad
\end{align}
as required.
 \bibliographystyle{utphys.bst}
\bibliography{dense.bib}

\providecommand{\href}[2]{#2}\begingroup\raggedright\begin{thebibliography}{10}

\bibitem{Bjorken:1982tu}
J.~D. Bjorken, ``{Energy Loss of Energetic Partons in Quark - Gluon Plasma:
  Possible Extinction of High p(t) Jets in Hadron - Hadron Collisions},''.

\bibitem{BRAHMS:2004adc}
{\bfseries BRAHMS} Collaboration, I.~Arsene {\em et~al.}, ``{Quark gluon plasma
  and color glass condensate at RHIC? The Perspective from the BRAHMS
  experiment},'' \href{http://dx.doi.org/10.1016/j.nuclphysa.2005.02.130}{{\em
  Nucl. Phys. A} {\bfseries 757} (2005) 1--27},
  \href{http://arxiv.org/abs/nucl-ex/0410020}{{\ttfamily
  arXiv:nucl-ex/0410020}}.

\bibitem{PHOBOS:2004zne}
{\bfseries PHOBOS} Collaboration, B.~B. Back {\em et~al.}, ``{The PHOBOS
  perspective on discoveries at RHIC},''
  \href{http://dx.doi.org/10.1016/j.nuclphysa.2005.03.084}{{\em Nucl. Phys. A}
  {\bfseries 757} (2005) 28--101},
  \href{http://arxiv.org/abs/nucl-ex/0410022}{{\ttfamily
  arXiv:nucl-ex/0410022}}.

\bibitem{STAR:2005gfr}
{\bfseries STAR} Collaboration, J.~Adams {\em et~al.}, ``{Experimental and
  theoretical challenges in the search for the quark gluon plasma: The STAR
  Collaboration's critical assessment of the evidence from RHIC collisions},''
  \href{http://dx.doi.org/10.1016/j.nuclphysa.2005.03.085}{{\em Nucl. Phys. A}
  {\bfseries 757} (2005) 102--183},
  \href{http://arxiv.org/abs/nucl-ex/0501009}{{\ttfamily
  arXiv:nucl-ex/0501009}}.

\bibitem{PHENIX:2004vcz}
{\bfseries PHENIX} Collaboration, K.~Adcox {\em et~al.}, ``{Formation of dense
  partonic matter in relativistic nucleus-nucleus collisions at RHIC:
  Experimental evaluation by the PHENIX collaboration},''
  \href{http://dx.doi.org/10.1016/j.nuclphysa.2005.03.086}{{\em Nucl. Phys. A}
  {\bfseries 757} (2005) 184--283},
  \href{http://arxiv.org/abs/nucl-ex/0410003}{{\ttfamily
  arXiv:nucl-ex/0410003}}.

\bibitem{ATLAS:2010isq}
{\bfseries ATLAS} Collaboration, G.~Aad {\em et~al.}, ``{Observation of a
  Centrality-Dependent Dijet Asymmetry in Lead-Lead Collisions at
  $\sqrt{s_{NN}}=2.77$ TeV with the ATLAS Detector at the LHC},''
  \href{http://dx.doi.org/10.1103/PhysRevLett.105.252303}{{\em Phys. Rev.
  Lett.} {\bfseries 105} (2010) 252303},
  \href{http://arxiv.org/abs/1011.6182}{{\ttfamily arXiv:1011.6182 [hep-ex]}}.

\bibitem{ALICE:2010yje}
{\bfseries ALICE} Collaboration, K.~Aamodt {\em et~al.}, ``{Suppression of
  Charged Particle Production at Large Transverse Momentum in Central Pb-Pb
  Collisions at $\sqrt{s_{NN}} =$ 2.76 TeV},''
  \href{http://dx.doi.org/10.1016/j.physletb.2010.12.020}{{\em Phys. Lett. B}
  {\bfseries 696} (2011) 30--39},
  \href{http://arxiv.org/abs/1012.1004}{{\ttfamily arXiv:1012.1004 [nucl-ex]}}.

\bibitem{CMS:2011iwn}
{\bfseries CMS} Collaboration, S.~Chatrchyan {\em et~al.}, ``{Observation and
  studies of jet quenching in PbPb collisions at nucleon-nucleon center-of-mass
  energy = 2.76 TeV},''
  \href{http://dx.doi.org/10.1103/PhysRevC.84.024906}{{\em Phys. Rev. C}
  {\bfseries 84} (2011) 024906},
  \href{http://arxiv.org/abs/1102.1957}{{\ttfamily arXiv:1102.1957 [nucl-ex]}}.

\bibitem{ATLAS:2018gwx}
{\bfseries ATLAS} Collaboration, M.~Aaboud {\em et~al.}, ``{Measurement of the
  nuclear modification factor for inclusive jets in Pb+Pb collisions at
  $\sqrt{s_\mathrm{NN}}=5.02$ TeV with the ATLAS detector},''
  \href{http://dx.doi.org/10.1016/j.physletb.2018.10.076}{{\em Phys. Lett. B}
  {\bfseries 790} (2019) 108--128},
  \href{http://arxiv.org/abs/1805.05635}{{\ttfamily arXiv:1805.05635
  [nucl-ex]}}.

\bibitem{CMS:2021vui}
{\bfseries CMS} Collaboration, A.~M. Sirunyan {\em et~al.}, ``{First
  measurement of large area jet transverse momentum spectra in heavy-ion
  collisions},'' \href{http://dx.doi.org/10.1007/JHEP05(2021)284}{{\em JHEP}
  {\bfseries 05} (2021) 284}, \href{http://arxiv.org/abs/2102.13080}{{\ttfamily
  arXiv:2102.13080 [hep-ex]}}.

\bibitem{ALICE:2023waz}
{\bfseries ALICE} Collaboration, S.~Acharya {\em et~al.}, ``{Measurement of the
  radius dependence of charged-particle jet suppression in Pb\textendash{}Pb
  collisions at sNN=5.02TeV},''
  \href{http://dx.doi.org/10.1016/j.physletb.2023.138412}{{\em Phys. Lett. B}
  {\bfseries 849} (2024) 138412},
  \href{http://arxiv.org/abs/2303.00592}{{\ttfamily arXiv:2303.00592
  [nucl-ex]}}.

\bibitem{Connors:2017ptx}
M.~Connors, C.~Nattrass, R.~Reed, and S.~Salur, ``{Jet measurements in heavy
  ion physics},'' \href{http://dx.doi.org/10.1103/RevModPhys.90.025005}{{\em
  Rev. Mod. Phys.} {\bfseries 90} (2018) 025005},
  \href{http://arxiv.org/abs/1705.01974}{{\ttfamily arXiv:1705.01974
  [nucl-ex]}}.

\bibitem{Wiedemann:2009sh}
U.~A. Wiedemann, ``{Jet Quenching in Heavy Ion Collisions},''
  \href{http://arxiv.org/abs/0908.2306}{{\ttfamily arXiv:0908.2306 [hep-ph]}}.

\bibitem{Ke:2024emw}
W.~Ke, J.~Terry, and I.~Vitev, ``{Anisotropic jet broadening and jet shape},''
  \href{http://arxiv.org/abs/2412.12250}{{\ttfamily arXiv:2412.12250
  [hep-ph]}}.

\bibitem{Chien:2024uax}
Y.-T. Chien, O.~Fedkevych, D.~Reichelt, and S.~Schumann, ``{Jet angularities in
  dijet production in proton-proton and heavy-ion collisions at RHIC},''
  \href{http://dx.doi.org/10.1007/JHEP07(2024)230}{{\em JHEP} {\bfseries 07}
  (2024) 230}, \href{http://arxiv.org/abs/2404.04168}{{\ttfamily
  arXiv:2404.04168 [hep-ph]}}.

\bibitem{Budhraja:2023rgo}
A.~Budhraja, R.~Sharma, and B.~Singh, ``{Medium modifications to jet
  angularities using SCET with Glauber gluons},''
  \href{http://arxiv.org/abs/2305.10237}{{\ttfamily arXiv:2305.10237
  [hep-ph]}}.

\bibitem{Barata:2024bmx}
J.~a. Barata, I.~Moult, and J.~a.~M. Silva, ``{Tracking Energy Loss in Heavy
  Ion Collisions},'' \href{http://arxiv.org/abs/2409.18174}{{\ttfamily
  arXiv:2409.18174 [hep-ph]}}.

\bibitem{Mehtar-Tani:2024mvl}
Y.~Mehtar-Tani, ``{Non-linear dynamics of jet quenching},''
  \href{http://arxiv.org/abs/2411.11992}{{\ttfamily arXiv:2411.11992
  [hep-ph]}}.

\bibitem{Barata:2024ieg}
J.~a. Barata, M.~V. Kuzmin, J.~G. Milhano, and A.~V. Sadofyev, ``{Jet EEC
  aWAKEning: hydrodynamic response on the celestial sphere},''
  \href{http://arxiv.org/abs/2412.03616}{{\ttfamily arXiv:2412.03616
  [hep-ph]}}.

\bibitem{Li:2024pfi}
Y.~Li, S.-Y. Chen, W.~Kong, S.~Wang, and B.-W. Zhang, ``{Medium modifications
  of heavy-flavor jet angularities in high-energy nuclear collisions},''
  \href{http://arxiv.org/abs/2409.12742}{{\ttfamily arXiv:2409.12742
  [hep-ph]}}.

\bibitem{Larkoski:2017jix}
A.~J. Larkoski, I.~Moult, and B.~Nachman, ``{Jet Substructure at the Large
  Hadron Collider: A Review of Recent Advances in Theory and Machine
  Learning},'' \href{http://dx.doi.org/10.1016/j.physrep.2019.11.001}{{\em
  Phys. Rept.} {\bfseries 841} (2020) 1--63},
  \href{http://arxiv.org/abs/1709.04464}{{\ttfamily arXiv:1709.04464
  [hep-ph]}}.

\bibitem{Asquith:2018igt}
R.~Kogler {\em et~al.}, ``{Jet Substructure at the Large Hadron Collider:
  Experimental Review},''
  \href{http://dx.doi.org/10.1103/RevModPhys.91.045003}{{\em Rev. Mod. Phys.}
  {\bfseries 91} no.~4, (2019) 045003},
  \href{http://arxiv.org/abs/1803.06991}{{\ttfamily arXiv:1803.06991
  [hep-ex]}}.

\bibitem{Marzani:2019hun}
S.~Marzani, G.~Soyez, and M.~Spannowsky,
  \href{http://dx.doi.org/10.1007/978-3-030-15709-8}{{\em {Looking inside jets:
  an introduction to jet substructure and boosted-object phenomenology}}},
  vol.~958.
\newblock Springer, 2019.
\newblock \href{http://arxiv.org/abs/1901.10342}{{\ttfamily arXiv:1901.10342
  [hep-ph]}}.

\bibitem{Collins:1989gx}
J.~C. Collins, D.~E. Soper, and G.~F. Sterman, ``{Factorization of Hard
  Processes in QCD},'' \href{http://dx.doi.org/10.1142/9789814503266_0001}{{\em
  Adv. Ser. Direct. High Energy Phys.} {\bfseries 5} (1989) 1--91},
  \href{http://arxiv.org/abs/hep-ph/0409313}{{\ttfamily arXiv:hep-ph/0409313}}.

\bibitem{Landau:1953um}
L.~D. Landau and I.~Pomeranchuk, ``{Limits of applicability of the theory of
  bremsstrahlung electrons and pair production at high-energies},'' {\em Dokl.
  Akad. Nauk Ser. Fiz.} {\bfseries 92} (1953) 535--536.

\bibitem{Migdal:1956tc}
A.~B. Migdal, ``{Bremsstrahlung and pair production in condensed media at
  high-energies},'' \href{http://dx.doi.org/10.1103/PhysRev.103.1811}{{\em
  Phys. Rev.} {\bfseries 103} (1956) 1811--1820}.

\bibitem{Gyulassy:1993hr}
M.~Gyulassy and X.-n. Wang, ``{Multiple collisions and induced gluon
  Bremsstrahlung in QCD},''
  \href{http://dx.doi.org/10.1016/0550-3213(94)90079-5}{{\em Nucl. Phys. B}
  {\bfseries 420} (1994) 583--614},
  \href{http://arxiv.org/abs/nucl-th/9306003}{{\ttfamily
  arXiv:nucl-th/9306003}}.

\bibitem{Wang:1994fx}
X.-N. Wang, M.~Gyulassy, and M.~Plumer, ``{The LPM effect in QCD and radiative
  energy loss in a quark gluon plasma},''
  \href{http://dx.doi.org/10.1103/PhysRevD.51.3436}{{\em Phys. Rev. D}
  {\bfseries 51} (1995) 3436--3446},
  \href{http://arxiv.org/abs/hep-ph/9408344}{{\ttfamily arXiv:hep-ph/9408344}}.

\bibitem{Baier:1994bd}
R.~Baier, Y.~L. Dokshitzer, S.~Peigne, and D.~Schiff, ``{Induced gluon
  radiation in a QCD medium},''
  \href{http://dx.doi.org/10.1016/0370-2693(94)01617-L}{{\em Phys. Lett. B}
  {\bfseries 345} (1995) 277--286},
  \href{http://arxiv.org/abs/hep-ph/9411409}{{\ttfamily arXiv:hep-ph/9411409}}.

\bibitem{Baier:1996kr}
R.~Baier, Y.~L. Dokshitzer, A.~H. Mueller, S.~Peigne, and D.~Schiff,
  ``{Radiative energy loss of high-energy quarks and gluons in a finite volume
  quark - gluon plasma},''
  \href{http://dx.doi.org/10.1016/S0550-3213(96)00553-6}{{\em Nucl. Phys. B}
  {\bfseries 483} (1997) 291--320},
  \href{http://arxiv.org/abs/hep-ph/9607355}{{\ttfamily arXiv:hep-ph/9607355}}.

\bibitem{Baier:1996sk}
R.~Baier, Y.~L. Dokshitzer, A.~H. Mueller, S.~Peigne, and D.~Schiff,
  ``{Radiative energy loss and p(T) broadening of high-energy partons in
  nuclei},'' \href{http://dx.doi.org/10.1016/S0550-3213(96)00581-0}{{\em Nucl.
  Phys. B} {\bfseries 484} (1997) 265--282},
  \href{http://arxiv.org/abs/hep-ph/9608322}{{\ttfamily arXiv:hep-ph/9608322}}.

\bibitem{Zakharov:1996fv}
B.~G. Zakharov, ``{Fully quantum treatment of the Landau-Pomeranchuk-Migdal
  effect in QED and QCD},'' \href{http://dx.doi.org/10.1134/1.567126}{{\em JETP
  Lett.} {\bfseries 63} (1996) 952--957},
  \href{http://arxiv.org/abs/hep-ph/9607440}{{\ttfamily arXiv:hep-ph/9607440}}.

\bibitem{Zakharov:1997uu}
B.~G. Zakharov, ``{Radiative energy loss of high-energy quarks in finite size
  nuclear matter and quark - gluon plasma},''
  \href{http://dx.doi.org/10.1134/1.567389}{{\em JETP Lett.} {\bfseries 65}
  (1997) 615--620}, \href{http://arxiv.org/abs/hep-ph/9704255}{{\ttfamily
  arXiv:hep-ph/9704255}}.

\bibitem{Gyulassy:2000er}
M.~Gyulassy, P.~Levai, and I.~Vitev, ``{Reaction operator approach to
  nonAbelian energy loss},''
  \href{http://dx.doi.org/10.1016/S0550-3213(00)00652-0}{{\em Nucl. Phys. B}
  {\bfseries 594} (2001) 371--419},
  \href{http://arxiv.org/abs/nucl-th/0006010}{{\ttfamily
  arXiv:nucl-th/0006010}}.

\bibitem{Wiedemann:2000za}
U.~A. Wiedemann, ``{Gluon radiation off hard quarks in a nuclear environment:
  Opacity expansion},''
  \href{http://dx.doi.org/10.1016/S0550-3213(00)00457-0}{{\em Nucl. Phys. B}
  {\bfseries 588} (2000) 303--344},
  \href{http://arxiv.org/abs/hep-ph/0005129}{{\ttfamily arXiv:hep-ph/0005129}}.

\bibitem{Guo:2000nz}
X.-f. Guo and X.-N. Wang, ``{Multiple scattering, parton energy loss and
  modified fragmentation functions in deeply inelastic e A scattering},''
  \href{http://dx.doi.org/10.1103/PhysRevLett.85.3591}{{\em Phys. Rev. Lett.}
  {\bfseries 85} (2000) 3591--3594},
  \href{http://arxiv.org/abs/hep-ph/0005044}{{\ttfamily arXiv:hep-ph/0005044}}.

\bibitem{Wang:2001ifa}
X.-N. Wang and X.-f. Guo, ``{Multiple parton scattering in nuclei: Parton
  energy loss},'' \href{http://dx.doi.org/10.1016/S0375-9474(01)01130-7}{{\em
  Nucl. Phys. A} {\bfseries 696} (2001) 788--832},
  \href{http://arxiv.org/abs/hep-ph/0102230}{{\ttfamily arXiv:hep-ph/0102230}}.

\bibitem{Arnold:2002ja}
P.~B. Arnold, G.~D. Moore, and L.~G. Yaffe, ``{Photon and gluon emission in
  relativistic plasmas},''
  \href{http://dx.doi.org/10.1088/1126-6708/2002/06/030}{{\em JHEP} {\bfseries
  06} (2002) 030}, \href{http://arxiv.org/abs/hep-ph/0204343}{{\ttfamily
  arXiv:hep-ph/0204343}}.

\bibitem{Arnold:2002zm}
P.~B. Arnold, G.~D. Moore, and L.~G. Yaffe, ``{Effective kinetic theory for
  high temperature gauge theories},''
  \href{http://dx.doi.org/10.1088/1126-6708/2003/01/030}{{\em JHEP} {\bfseries
  01} (2003) 030}, \href{http://arxiv.org/abs/hep-ph/0209353}{{\ttfamily
  arXiv:hep-ph/0209353}}.

\bibitem{Salgado:2003gb}
C.~A. Salgado and U.~A. Wiedemann, ``{Calculating quenching weights},''
  \href{http://dx.doi.org/10.1103/PhysRevD.68.014008}{{\em Phys. Rev. D}
  {\bfseries 68} (2003) 014008},
  \href{http://arxiv.org/abs/hep-ph/0302184}{{\ttfamily arXiv:hep-ph/0302184}}.

\bibitem{Liu:2006ug}
H.~Liu, K.~Rajagopal, and U.~A. Wiedemann, ``{Calculating the jet quenching
  parameter from AdS/CFT},''
  \href{http://dx.doi.org/10.1103/PhysRevLett.97.182301}{{\em Phys. Rev. Lett.}
  {\bfseries 97} (2006) 182301},
  \href{http://arxiv.org/abs/hep-ph/0605178}{{\ttfamily arXiv:hep-ph/0605178}}.

\bibitem{Qin:2007rn}
G.-Y. Qin, J.~Ruppert, C.~Gale, S.~Jeon, G.~D. Moore, and M.~G. Mustafa,
  ``{Radiative and collisional jet energy loss in the quark-gluon plasma at
  RHIC},'' \href{http://dx.doi.org/10.1103/PhysRevLett.100.072301}{{\em Phys.
  Rev. Lett.} {\bfseries 100} (2008) 072301},
  \href{http://arxiv.org/abs/0710.0605}{{\ttfamily arXiv:0710.0605 [hep-ph]}}.

\bibitem{Armesto:2011ht}
N.~Armesto {\em et~al.}, ``{Comparison of Jet Quenching Formalisms for a
  Quark-Gluon Plasma 'Brick'},''
  \href{http://dx.doi.org/10.1103/PhysRevC.86.064904}{{\em Phys. Rev. C}
  {\bfseries 86} (2012) 064904},
  \href{http://arxiv.org/abs/1106.1106}{{\ttfamily arXiv:1106.1106 [hep-ph]}}.

\bibitem{Mehtar-Tani:2010ebp}
Y.~Mehtar-Tani, C.~A. Salgado, and K.~Tywoniuk, ``{Anti-angular ordering of
  gluon radiation in QCD media},''
  \href{http://dx.doi.org/10.1103/PhysRevLett.106.122002}{{\em Phys. Rev.
  Lett.} {\bfseries 106} (2011) 122002},
  \href{http://arxiv.org/abs/1009.2965}{{\ttfamily arXiv:1009.2965 [hep-ph]}}.

\bibitem{Mehtar-Tani:2012mfa}
Y.~Mehtar-Tani, C.~A. Salgado, and K.~Tywoniuk, ``{The Radiation pattern of a
  QCD antenna in a dense medium},''
  \href{http://dx.doi.org/10.1007/JHEP10(2012)197}{{\em JHEP} {\bfseries 10}
  (2012) 197}, \href{http://arxiv.org/abs/1205.5739}{{\ttfamily arXiv:1205.5739
  [hep-ph]}}.

\bibitem{Casalderrey-Solana:2011ule}
J.~Casalderrey-Solana and E.~Iancu, ``{Interference effects in medium-induced
  gluon radiation},'' \href{http://dx.doi.org/10.1007/JHEP08(2011)015}{{\em
  JHEP} {\bfseries 08} (2011) 015},
  \href{http://arxiv.org/abs/1105.1760}{{\ttfamily arXiv:1105.1760 [hep-ph]}}.

\bibitem{Casalderrey-Solana:2012evi}
J.~Casalderrey-Solana, Y.~Mehtar-Tani, C.~A. Salgado, and K.~Tywoniuk, ``{New
  picture of jet quenching dictated by color coherence},''
  \href{http://dx.doi.org/10.1016/j.physletb.2013.07.046}{{\em Phys. Lett. B}
  {\bfseries 725} (2013) 357--360},
  \href{http://arxiv.org/abs/1210.7765}{{\ttfamily arXiv:1210.7765 [hep-ph]}}.

\bibitem{antenna_dense}
Y.~Mehtar-Tani, C.~A. Salgado, and K.~Tywoniuk, ``{The Radiation pattern of a
  QCD antenna in a dense medium},''
  \href{http://dx.doi.org/10.1007/JHEP10(2012)197}{{\em JHEP} {\bfseries 10}
  (2012) 197}, \href{http://arxiv.org/abs/1205.5739}{{\ttfamily arXiv:1205.5739
  [hep-ph]}}.

\bibitem{Mehtar-Tani:2024smp}
Y.~Mehtar-Tani, F.~Ringer, B.~Singh, and V.~Vaidya, ``{Factorization for jet
  production in heavy-ion collisions},''
  \href{http://arxiv.org/abs/2409.05957}{{\ttfamily arXiv:2409.05957
  [hep-ph]}}.

\bibitem{Singh:2024vwb}
B.~Singh and V.~Vaidya, ``{Factorization for energy-energy correlator in heavy
  ion collision},'' \href{http://arxiv.org/abs/2408.02753}{{\ttfamily
  arXiv:2408.02753 [hep-ph]}}.

\bibitem{Bauer:2000yr}
C.~W. Bauer, S.~Fleming, D.~Pirjol, and I.~W. Stewart, ``{An Effective field
  theory for collinear and soft gluons: Heavy to light decays},''
  \href{http://dx.doi.org/10.1103/PhysRevD.63.114020}{{\em Phys. Rev. D}
  {\bfseries 63} (2001) 114020},
  \href{http://arxiv.org/abs/hep-ph/0011336}{{\ttfamily arXiv:hep-ph/0011336}}.

\bibitem{Bauer:2002nz}
C.~W. Bauer, S.~Fleming, D.~Pirjol, I.~Z. Rothstein, and I.~W. Stewart, ``{Hard
  scattering factorization from effective field theory},''
  \href{http://dx.doi.org/10.1103/PhysRevD.66.014017}{{\em Phys. Rev. D}
  {\bfseries 66} (2002) 014017},
  \href{http://arxiv.org/abs/hep-ph/0202088}{{\ttfamily arXiv:hep-ph/0202088}}.

\bibitem{Rothstein:2016bsq}
I.~Z. Rothstein and I.~W. Stewart, ``{An Effective Field Theory for Forward
  Scattering and Factorization Violation},''
  \href{http://dx.doi.org/10.1007/JHEP08(2016)025}{{\em JHEP} {\bfseries 08}
  (2016) 025}, \href{http://arxiv.org/abs/1601.04695}{{\ttfamily
  arXiv:1601.04695 [hep-ph]}}.

\bibitem{Dasgupta:2014yra}
M.~Dasgupta, F.~Dreyer, G.~P. Salam, and G.~Soyez, ``{Small-radius jets to all
  orders in QCD},'' \href{http://dx.doi.org/10.1007/JHEP04(2015)039}{{\em JHEP}
  {\bfseries 04} (2015) 039}, \href{http://arxiv.org/abs/1411.5182}{{\ttfamily
  arXiv:1411.5182 [hep-ph]}}.

\bibitem{Kaufmann:2015hma}
T.~Kaufmann, A.~Mukherjee, and W.~Vogelsang, ``{Hadron Fragmentation Inside
  Jets in Hadronic Collisions},''
  \href{http://dx.doi.org/10.1103/PhysRevD.92.054015}{{\em Phys. Rev. D}
  {\bfseries 92} no.~5, (2015) 054015},
  \href{http://arxiv.org/abs/1506.01415}{{\ttfamily arXiv:1506.01415
  [hep-ph]}}. [Erratum: Phys.Rev.D 101, 079901 (2020)].

\bibitem{Kang:2016mcy}
Z.-B. Kang, F.~Ringer, and I.~Vitev, ``{The semi-inclusive jet function in SCET
  and small radius resummation for inclusive jet production},''
  \href{http://dx.doi.org/10.1007/JHEP10(2016)125}{{\em JHEP} {\bfseries 10}
  (2016) 125}, \href{http://arxiv.org/abs/1606.06732}{{\ttfamily
  arXiv:1606.06732 [hep-ph]}}.

\bibitem{Dai:2016hzf}
L.~Dai, C.~Kim, and A.~K. Leibovich, ``{Fragmentation of a Jet with Small
  Radius},'' \href{http://dx.doi.org/10.1103/PhysRevD.94.114023}{{\em Phys.
  Rev. D} {\bfseries 94} no.~11, (2016) 114023},
  \href{http://arxiv.org/abs/1606.07411}{{\ttfamily arXiv:1606.07411
  [hep-ph]}}.

\bibitem{vanBeekveld:2024jnx}
M.~van Beekveld, M.~Dasgupta, B.~K. El-Menoufi, J.~Helliwell, A.~Karlberg, and
  P.~F. Monni, ``{Two-loop anomalous dimensions for small-R jet versus hadronic
  fragmentation functions},''
  \href{http://dx.doi.org/10.1007/JHEP07(2024)239}{{\em JHEP} {\bfseries 07}
  (2024) 239}, \href{http://arxiv.org/abs/2402.05170}{{\ttfamily
  arXiv:2402.05170 [hep-ph]}}.

\bibitem{Lee:2024icn}
K.~Lee, I.~Moult, and X.~Zhang, ``{Revisiting Single Inclusive Jet Production:
  Timelike Factorization and Reciprocity},''
  \href{http://arxiv.org/abs/2409.19045}{{\ttfamily arXiv:2409.19045
  [hep-ph]}}.

\bibitem{Dasgupta:2001sh}
M.~Dasgupta and G.~P. Salam, ``{Resummation of nonglobal QCD observables},''
  \href{http://dx.doi.org/10.1016/S0370-2693(01)00725-0}{{\em Phys. Lett. B}
  {\bfseries 512} (2001) 323--330},
  \href{http://arxiv.org/abs/hep-ph/0104277}{{\ttfamily arXiv:hep-ph/0104277}}.

\bibitem{Larkoski:2015zka}
A.~J. Larkoski, I.~Moult, and D.~Neill, ``{Non-Global Logarithms,
  Factorization, and the Soft Substructure of Jets},''
  \href{http://dx.doi.org/10.1007/JHEP09(2015)143}{{\em JHEP} {\bfseries 09}
  (2015) 143}, \href{http://arxiv.org/abs/1501.04596}{{\ttfamily
  arXiv:1501.04596 [hep-ph]}}.

\bibitem{Becher:2015hka}
T.~Becher, M.~Neubert, L.~Rothen, and D.~Y. Shao, ``{Effective Field Theory for
  Jet Processes},''
  \href{http://dx.doi.org/10.1103/PhysRevLett.116.192001}{{\em Phys. Rev.
  Lett.} {\bfseries 116} no.~19, (2016) 192001},
  \href{http://arxiv.org/abs/1508.06645}{{\ttfamily arXiv:1508.06645
  [hep-ph]}}.

\bibitem{Gyulassy:2000fs}
M.~Gyulassy, P.~Levai, and I.~Vitev, ``{NonAbelian energy loss at finite
  opacity},'' \href{http://dx.doi.org/10.1103/PhysRevLett.85.5535}{{\em Phys.
  Rev. Lett.} {\bfseries 85} (2000) 5535--5538},
  \href{http://arxiv.org/abs/nucl-th/0005032}{{\ttfamily
  arXiv:nucl-th/0005032}}.

\bibitem{Kumar:2019uvu}
A.~Kumar, A.~Majumder, and C.~Shen, ``{Energy and scale dependence of $\hat{q}$
  and the \textquotedblleft{}JET puzzle\textquotedblright{}},''
  \href{http://dx.doi.org/10.1103/PhysRevC.101.034908}{{\em Phys. Rev. C}
  {\bfseries 101} no.~3, (2020) 034908},
  \href{http://arxiv.org/abs/1909.03178}{{\ttfamily arXiv:1909.03178
  [nucl-th]}}.

\bibitem{DEramo:2018eoy}
F.~D'Eramo, K.~Rajagopal, and Y.~Yin, ``{Moli\`ere scattering in quark-gluon
  plasma: finding point-like scatterers in a liquid},''
  \href{http://dx.doi.org/10.1007/JHEP01(2019)172}{{\em JHEP} {\bfseries 01}
  (2019) 172}, \href{http://arxiv.org/abs/1808.03250}{{\ttfamily
  arXiv:1808.03250 [hep-ph]}}.

\end{thebibliography}\endgroup
\end{document}